\documentclass{pasj01}

\usepackage{graphicx} 
\usepackage{natbib}

\begin{document} 
\Received{}
\Accepted{}

\title{Dust evolution processes constrained by extinction curves in nearby galaxies}

\author{Kuan-Chou \textsc{Hou}\altaffilmark{1,2}%
}
\altaffiltext{1}{Institute of Astronomy, and Astrophysics, Academia Sinica, PO Box 23-141, Taipei 10617, Taiwan}
\altaffiltext{2}{Department of Physics, Institute of Astrophysics, National Taiwan University, Taipei 10617, Taiwan}
\email{kchou@asiaa.sinica.edu.tw}

\author{Hiroyuki \textsc{Hirashita}\altaffilmark{1}}
\author{Micha{\l} J. \textsc{Micha{\l}owski}\altaffilmark{3}}
\altaffiltext{3}{SUPA, Institute for Astronomy, University of Edinburgh, Royal Observatory, Blackford Hill, Edinburgh, EH9 3HJ, UK}



\maketitle

\begin{abstract}
Extinction curves, especially those
in the Milky Way (MW), the Large Magellanic Cloud (LMC), and
the Small Magellanic Cloud (SMC), have provided us with a clue to the
dust properties in the nearby Universe.
We examine whether or not these extinction curves can be explained
by well known dust evolution processes.
We treat the dust production in stellar ejecta, destruction in supernova shocks, dust growth by accretion and coagulation, and dust disruption by shattering.
To make a survey of the large parameter space possible,
we simplify the treatment of the grain size distribution evolution by
adopting the `two-size approximation', in which
we divide the grain population into
small ($\lesssim 0.03~\mu$m) and large ($\gtrsim 0.03~\mu$m) grains.
It is confirmed that the MW extinction curve can be reproduced 
in reasonable ranges for the time-scale of the above processes 
with a silicate-graphite mixture. This indicates that the MW
extinction curve is a natural consequence of the dust evolution through
the above processes.
We also find that the same models fail to reproduce the SMC/LMC extinction
curves. Nevertheless, this failure can be remedied by
giving higher supernova destruction rates
for small carbonaceous dust
and considering amorphous carbon for carbonaceous dust;
these modification fall in fact in line with previous studies.
Therefore, we conclude that the current dust evolution scenario
composed of the aforementioned processes is successful in
explaining the extinction curves.
All the extinction curves favor efficient interstellar processing of dust,
especially, strong grain growth by accretion and coagulation.
\end{abstract}

\section{Introduction}
Dust enrichment is one of the most important aspects
for understanding the evolution of galaxies.
Dust absorbs and scatters the stellar light and reemits it
in the far infrared, thereby
shaping the spectral energy distribution (SED)
\citep[e.g.][for recent modeling]{Yajima:2014aa,Schaerer:2015aa}.
The surface of dust grains provides a condition suitable for 
efficient formation of molecular hydrogen, which is an important 
coolant in low-metallicity clouds 
\citep[e.g.][]{Cazaux:2004aa}.
Dust itself is also an important coolant in star formation,
inducing the final fragmentation to determine the stellar mass 
\citep{Omukai:2005aa,Schneider:2006aa}.

For the important role of dust in determining the SED of
galaxies, the wavelength dependence of extinction
(extinction is the sum of absorption and scattering),
the so-called extinction curve, is the central quantity.
Moreover, extinction curves reflect
grain size distribution and grain composition
\citep[e.g.][]{Weingartner:2001aa}
, both of
which are important in determining the aforementioned processes,
i.e.\ grain surface
reactions and dust cooling \citep{Yamasawa:2011aa}.

Because of the proximity, the most detailed studies
of extinction curves have been performed in
the Milky Way (MW), the Large Magellanic Cloud (LMC), and
the Small Magellanic Cloud (SMC), where the foreground
extinction for each
individual bright stellar source can be measured
\citep[e.g.][]{Pei:1992aa,Gordon:2003aa}. 
There have been various theoretical
efforts of explaining these extinction curves through the
modeling of grain composition and size distribution.
\citet*[][hereafter MRN]{Mathis:1977aa} 
showed that a mixture of silicate
and graphite with a grain size distribution of a
power law $\propto a^{-3.5}$ ($a$ is the grain radius
and they considered ($a\sim $0.005--0.25 $\micron$)
reproduces the MW extinction curve
\citep*[see also][]{Draine:1984aa,Kim:1994aa}, 
\citet{Pei:1992aa}
showed that the extinction curves in the LMC and SMC
are also explained by the same grain size distribution
with different abundance ratios between silicate and
graphite. \citet{Weingartner:2001aa} applied
more detailed functional forms of grain size distributions
to explain the extinction curves in the MW, LMC, and SMC.
Although these models are successful in explaining the
extinction curves, we still need to clarify how
such grain size distributions as assumed in their models
are established as a consequence of dust evolution in galaxies.

There have been some efforts of modeling the evolution of
grain size distribution in galaxies.
\citet{Liffman:1989aa}
treated the evolution of grain size distribution by
considering the formation of refractory dust in stellar
ejecta, the growth of mantles in the dense medium, and
dust destruction in supernova shocks. Although their
efforts of treating the grain size distribution is pioneering,
their method based on tracing individual particles cannot
treat grain fragmentation (or shattering) and grain sticking
(or coagulation).
\citet{ODonnell:1997aa} incorporated coagulation, shattering,
and shock destruction for the dust processing
mechanisms and roughly reproduced
the MW extinction curve
by the grain size distribution calculated in their models.
Their results indicate the importance of these grain
processing mechanisms in explaining the extinction curve.
However, they only discussed the condition specific for the
current MW, and the question still remains as to whether or not
the MW extinction curve is reproduced as a result of dust
evolution, which is coupled with the dust enrichment history
of the MW.
Moreover, the different shapes of extinction curves in the LMC
and SMC still need to be addressed.

Recently, the evolution of grain size distribution has
been formulated in a consistent manner with galaxy evolution.
\citet{Asano:2013aa} constructed a framework for treating
the evolution of grain size distribution over the entire galaxy 
history. In their calculation, dust formed in stellar ejecta,
that is,
supernovae (SNe) and asymptotic giant branch (AGB) star winds,
dominate the grain size distribution at the early stage of galactic
evolution \citep{Bianchi:2007aa,Nozawa:2007aa,Valiante:2009aa,Gall:2011aa,Yasuda:2012aa}.
\citet{Asano:2013aa} assumed that these stellar sources form large ($\sim 0.1~\micron$)
grains, based on theoretical and observational evidence
(see section 2.1 of H15 and Section 2.1 of the present paper).
Thus, the dust is dominated by large grains at the early stage of 
galaxy evolution. As the system is enriched with dust,
shattering as a result of grain--grain collision becomes efficient
enough to increase the abundance of small grains.
The increase of small grains drastically enhances the
total grain surface area; as a consequence, grain growth by
the accretion of gas-phase metals becomes the most 
important process for dust enrichment. Afterwards, the
abundant small grains coagulate to form large grains. 
Throughout the galaxy evolution, dust destruction by 
SN shocks in the ISM is the main loss mechanism of dust mass 
\citep[see also][]{Dwek:1980aa}.
\citet{Asano:2014aa} calculated the evolution of extinction
curve based on their model of grain size distribution.
Their results tend to predict steeper extinction curves
than the MW curve because grain growth by accretion
drastically increases the abundance of small grains.
\citet{Nozawa:2015aa} showed, however, that the MW
extinction curve is reproduced by considering dense
molecular clouds in which strong coagulation converts
small grains to large grains
and flattens the extinction curve. Their model also
explains the extinction curves observed in high-redshift
quasars.

Although the above recent models took into account the
full details of grain formation and processing mechanisms,
there are still some uncertain free parameters regarding the time-scales
of individual processes. In particular, the time-scales (or efficiencies) of
accretion, shattering, and coagulation are strongly affected by
the density structures in the ISM, since
accretion and coagulation work only in the dense and cold ISM
while shattering occurs predominantly in the diffuse ISM
\citep{Asano:2013aa}. Therefore, to complement their detailed
models, a parameter survey study is desirable; that is,
we need to survey all the reasonable ranges
of the time-scales, for the purpose of checking if their
conclusions are sensitive to the assumed time-scales
or for the purpose of finding the ranges of the time-scales
that reproduce successfully the observed extinction curves.
However, their model based on a full treatment of grain
size distribution requires a lot of computational time,
and is not suitable for such a parameter survey study.

To make a parameter survey possible in a reasonable computational
time, we adopt a simplified model developed by
\citet[][hereafter H15]{Hirashita:2015aa} 
to calculate the
evolution of grain size distribution: they adopted a `two-size
approximation' approach, in which the grain sizes are
represented by two sizes: large ($\gtrsim 0.03~\mu$m) and small ($\lesssim 0.03~\mu$m) grains.
The model includes all the above processes considered by
\citet{Asano:2013aa} but only treat the mass exchange between
the small and large grain populations. H15 showed that
this two-size approximation traces the same evolutionary
behaviors of grain size distribution and extinction curve
as presented in
\citet{Asano:2013aa} and \citet{Asano:2014aa}. Therefore,
H15 concluded that the two-size approximation can be used
as a simplified (or computationally cheap) version of
the full treatment of grain size distribution.

Using this two-size approximation with two dust species,
silicate an graphite,
\citet{Bekki:2015aa} investigated the SMC/LMC extinction curves
with a framework of a one-zone chemical evolution.
They proposed a scenario in which small graphite grains
are transported out of the galaxy in the latest starburst
in the SMC (probably by radiation pressure), reproducing
the SMC extinction curve, which does not show a
2175 \AA\ bump caused by small graphite grains. This
scenario that small graphite grains are selectively lost by
dust wind reproduces the LMC extinction curve as well.
However, it is yet to be proven that
radiation pressure is effective in selective transport of small carbonaceous dust.

Although the above dust wind scenario can reproduce
the extinction curves, it is still worth considering if
dust processing mechanisms that work within a galaxy
explain the extinction curves. In other words, how well
the interstellar dust processing could reproduce the
MW, LMC, and SMC extinction curves is still to be
clarified.
In this paper, therefore, we use the dust evolution model based
on the two-size method developed in H15
and survey reasonable ranges of parameters characterizing
the time-scales of individual dust processing mechanisms.
We compare the extinction curves calculated by our models with
those observed in the MW, LMC and SMC, and examine
if these observed extinction curves can be reproduced by
the models. In addition, we will be able to constrain
the time-scales of dust processing mechanisms that govern the grain size distribution.
The simplicity of the two-size approach enables us to fully
survey the reasonable ranges of individual time-scales for the
first time. We also discuss a possibility that the
extinction curves of these three galaxies are simultaneously
explained with a single evolutionary scenario of dust
evolution.

This paper is organized as follows.
In Section 2, we describe the dust enrichment model and 
the calculation method of extinction curves. In Section 3, we show
the calculated extinction curves, which are compared with
the observed extinction curves in the MW, LMC, and SMC. We discuss
the results, laying particular emphasis on the time-scales
of various dust processing mechanisms. Finally, we
provide the conclusions 
in Section 5.

\section{Dust enrichment model}
We use the two-size dust enrichment model developed by H15. In this model,
the dust grains are divided into small and large grains, considering
that various grain processing mechanisms work differently between these two
grain populations. H15 proposed $a=0.03~\micron$, where $a$ is the grain radius,
for the boundary between the two populations based on the full grain
size calculations by \citet{Asano:2013aa}, whose models have been
successful in explaining the extinction curves in the MW and high-redshift
galaxies \citep{Nozawa:2015aa}. The model takes into account dust supply by stellar ejecta, dust destruction in supernova shocks, grain growth by accretion and coagulation and grain disruption by shattering. 
As formulated in \citet{Bekki:2015aa}, we also separately solve
silicate and carbonaceous dust. These two species are often adopted
to explain extinction curves \citep[e.g.][]{Weingartner:2001aa}.
Since the model successfully explained the dust abundance in
nearby galaxies in H15, we concentrate on extinction curves, which reflect
grain size distributions, in this paper.
Below we explain the models.
We adopt $Z_{\solar}$ = 0.02 for the solar metallicity throughout this
paper following H15.

\subsection{Two-size, two-species model}\label{subsec:twosize}

In H15, different dust species (i.e.\ silicate and carbonaceous dust) were not treated separately. In this paper, since dust material properties are important in reproducing extinction curves, we solve the evolution of different species, silicate and carbonaceous dust, separately.
We represent silicate and carbonaceous dust by the evolution of
Si and C, and assume that the mass fraction of Si in silicate is 0.166 while
that of C in carbonaceous dust is 1.
After applying the instantaneous recycling approximation \citep{Tinsley:1980aa},
the dust enrichment equations of each species for large and small grains are
written as (see H15 for the derivation)
\begin{equation}
  \mathcal{Y}_\mathrm{X}\frac{\mathrm{d}\mathcal{D}_\mathrm{l,X}}{\mathrm{d}Z} = f_\mathrm{in,X}(\mathcal{R}Z+\mathcal{Y}_\mathrm{X})+\beta_\mathrm{co}\mathcal{D}_\mathrm{s,X}-(\beta_\mathrm{SN}+\beta_\mathrm{sh}+\mathcal{R})\mathcal{D}_\mathrm{l,X},
\label{eq:large}
\end{equation}
\begin{equation}
  \mathcal{Y}_\mathrm{X}\frac{\mathrm{d}\mathcal{D}_\mathrm{s,X}}{\mathrm{d}Z} = \beta_\mathrm{sh}\mathcal{D}_\mathrm{l,X}-\left(\frac{\beta_\mathrm{SN}}{\alpha_\mathrm{X}}+\beta_\mathrm{co}+\mathcal{R}-\beta_\mathrm{acc}\right)\mathcal{D}_\mathrm{s,X},
\label{eq:small}
\end{equation}
where subscript X indicates the dust species (Si and C),
$\mathcal{D}_\mathrm{s,X}$ and $\mathcal{D}_\mathrm{l,X}$
are the dust-to-gas ratios of small grains and large grains, respectively,
$f_\mathrm{in,X}$ is the dust condensation efficiency of element X in the stellar ejecta,
$\mathcal{R}$ is the returned mass fraction from the formed stars,
$\mathcal{Y}_\mathrm{X}$ is the mass fraction of newly produced element X,
$\alpha_\mathrm{X}$ is the enhancement factor of SN destruction for small grains
relative to large grains,
and $\beta_\mathrm{SN}$, $\beta_\mathrm{sh}$, $\beta_\mathrm{co}$ and 
$\beta_\mathrm{acc}$ indicate the efficiencies (explained below) of 
supernova destruction, shattering, coagulation and accretion, respectively.
Note that these efficiencies ($\beta$s) depend on
$\mathcal{D}_{i,\mathrm{X}}$ ($i$ = s or l) except
$\beta_\mathrm{SN}$ (see below). Thus, $\beta$s depend on
material X, although we do not express this dependence explicitly for
the brevity of notation.

Now we explain equations (\ref{eq:large}) and (\ref{eq:small})
briefly. Equation (\ref{eq:large}) describes the increase of
large grains. The right-hand size of this equation represents
the stellar dust production
[$f_\mathrm{in,X}(\mathcal{R}Z+\mathcal{Y}_\mathrm{X})$],
the increase by coagulation
of small grains ($\beta_\mathrm{co}\mathcal{D}_\mathrm{s,X}$),
the decreases by SN destruction and shattering
($\beta_\mathrm{SN}\mathcal{D}_\mathrm{l,X}$ and
$\beta_\mathrm{sh}\mathcal{D}_\mathrm{l,X}$, respectively),
and the dilution of dust-to-gas ratio by returned gas
from stars ($\mathcal{RD}_\mathrm{l,X}$). Equation 
(\ref{eq:small}) shows the increase of small grains, and
it includes similar terms to equation (\ref{eq:large}).
Note that the shattering and coagulation terms in
equation (\ref{eq:small}) have
the opposite sign to those in equation (\ref{eq:large}),
since shattering is source and coagulation is sink for
small grains. The increase of dust abundance by accretion
($\beta_\mathrm{acc}\mathcal{D}_\mathrm{s,X}$) only appears
in equation (\ref{eq:small}); indeed, grain
growth by accretion is much more efficient for small grains than 
for large grains because small grains have much larger
surface-to-volume ratio (\citealt{Hirashita:2011aa,Asano:2013aa};
H15).

The reason why we assume that stellar ejecta provide large grains
in both SNe and AGB star winds is based on several theoretical and observational studies.
For observational evidence, 
\citet{Gall:2014aa} obtained the infrared spectrum
of SN 2010jl, suggesting that newly formed dust in SN ejecta
is dominated by large grains.
\citet{Scicluna:2015ab} showed with imaging polarimetry at optical and
near-infrared wavelengths that the
grains radii in the wind of a red supergiant star VY Canis Majoris is on
average $\sim 0.5~\micron$.
The typical size of grains produced by AGB stars is also suggested to be large ($a > 0.1 \mu$m) from observations of SEDs \citep{Groenewegen:1997aa,Gauger:1999aa}, and also
the polarization observation done by \citet{Norris:2012aa} 
supports large sizes of dust grains produced by AGB stars.
Several theoretical studies showed that reverse shocks
more efficiently destroy small grains than large grains,
which makes dust grains produced by SNe biased to large sizes
(\citealt{Nozawa:2007aa}; see also \citealt{Bianchi:2007aa}; \citealt{Marassi:2015aa}).
Theoretical studies have also shown that dust grains formed in AGB star winds
have large sizes \citep{Yasuda:2012aa,Ventura:2012aa}.
There is also evidence from meteoritic samples that dust species such as
SiC and graphite, whose isotopic compositions indicate AGB star origin,
have large grain sizes ($a\gtrsim 0.1~\micron$), supporting the formation of large
grains in AGB stars (e.g., Amari, Lewis, \& Anders 1994; Hoppe \& Zinner 2000).
These results robustly indicate that the sizes of grains originating from stellar sources
are biased to large sizes compared with the ISM grains;
as long as this is true, it is enough to include
stellar dust production only in equation (\ref{eq:large}) for the 
purpose of this paper. Moreover, if interstellar processing is more important
than stellar sources, the detailed assumption on the grain size distribution of
stellar dust is not essential.

We evaluate $\beta$s as follows:
$\beta_\mathrm{SN}\equiv\tau_\mathrm{SF}/\tau_\mathrm{SN}$,
$\beta_\mathrm{sh} \equiv \tau_\mathrm{SF}/ \tau_\mathrm{sh}$ and $\beta_\mathrm{co} \equiv \tau_\mathrm{SF}/ \tau_\mathrm{co}$,
where $\tau_\mathrm{SF}\equiv M_\mathrm{gas}/\psi$
($M_\mathrm{gas}$ is the total gas mass in the galaxy and $\psi$ is
the star formation rate) is the star formation time-scale,
and the shattering and coagulation time-scales are written as
\begin{equation}
  \tau_\mathrm{sh} = \tau_\mathrm{sh,0}\left(\frac{\mathcal{D}_\mathrm{l,X}}{\mathcal{D}_\mathrm{MW,l}}\right)^{-1},
\label{eq:sha}
\end{equation}
and
\begin{equation}
  \tau_\mathrm{co} = \tau_\mathrm{co,0}\left(\frac{\mathcal{D}_\mathrm{s,X}}{\mathcal{D}_\mathrm{MW,s}}\right)^{-1}, 
\label{eq:co}
\end{equation}
where the time-scales are normalized to $\tau_\mathrm{sh,0}$ and 
$\tau_\mathrm{co,0}$ at the MW dust-to-gas ratio, 
$\mathcal{D}_\mathrm{MW,l} = 0.007$  and $\mathcal{D}_\mathrm{MW,s} = 0.003$ 
(H15). Since shattering and coagulation are collisional processes,
their time-scales are inversely proportional to the dust-to-gas
ratio of the relevant species. The accretion efficiency
$\beta_\mathrm{acc}$ is regulated by the lifetime of dense
clouds ($\tau_\mathrm{cl}$), which host grain growth by accretion:
accretion is more efficient if $\tau_\mathrm{cl}$ is longer
and metallicity is higher. In the 
calculation of $\beta_\mathrm{acc}$, we also need atomic mass $m_\mathrm{X}$,
metal abundance $(\mathrm{X/H})$ normalized to solar 
abundance [$(\mathrm{X/H})_{\solar}$], mass fraction of key species $f_\mathrm{X}$ as given in
Table \ref{table:Adopted_quantities} 
(\citealt{Hirashita:2011aa}; H15).
We adopt $\alpha_\mathrm{X} = 1$ unless otherwise stated (we will vary
$\alpha_\mathrm{X}$ for the SMC and LMC later).

\begin{figure}
\begin{center}
\includegraphics[width=0.45\textwidth]{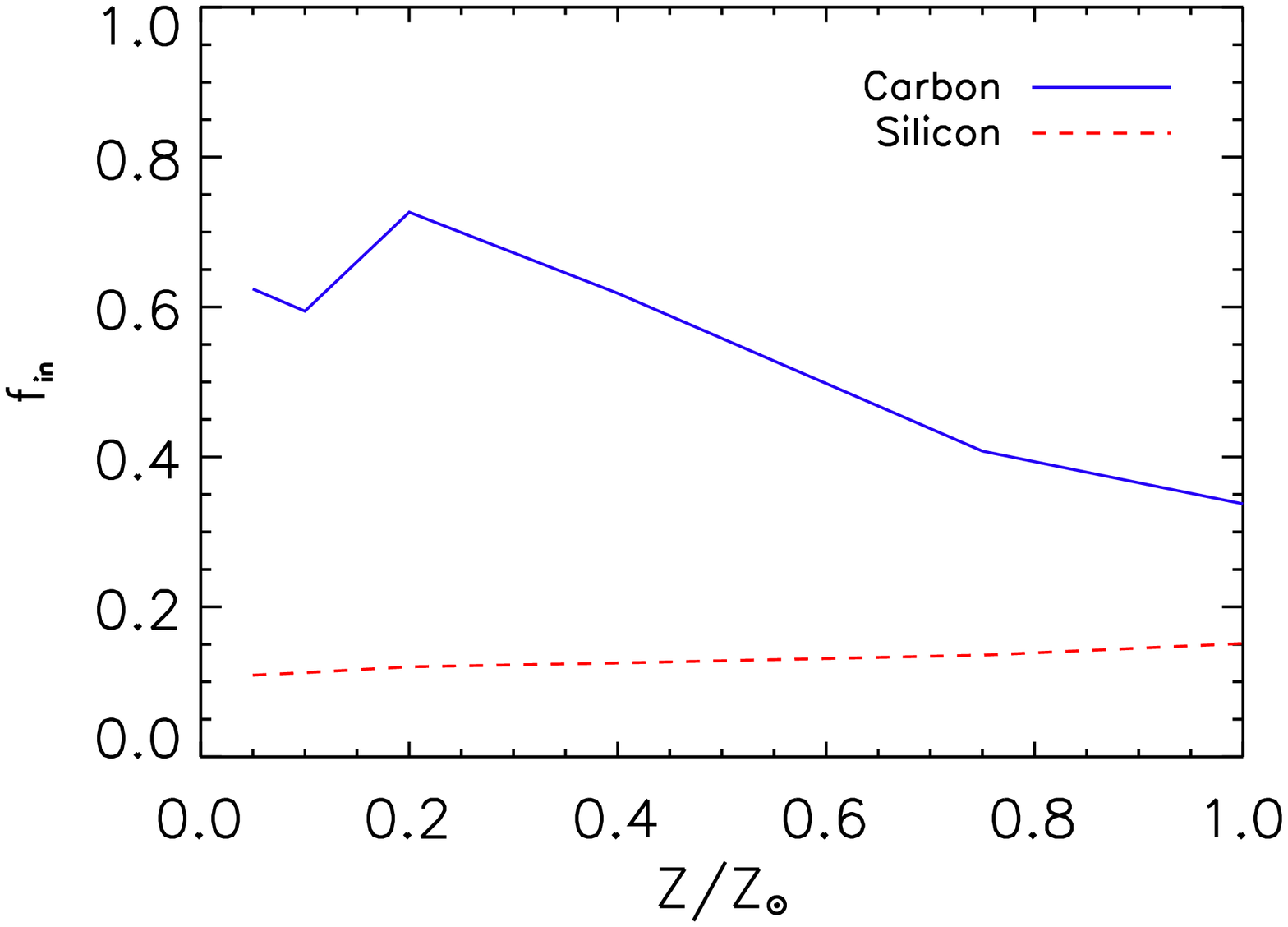}
\includegraphics[width=0.45\textwidth]{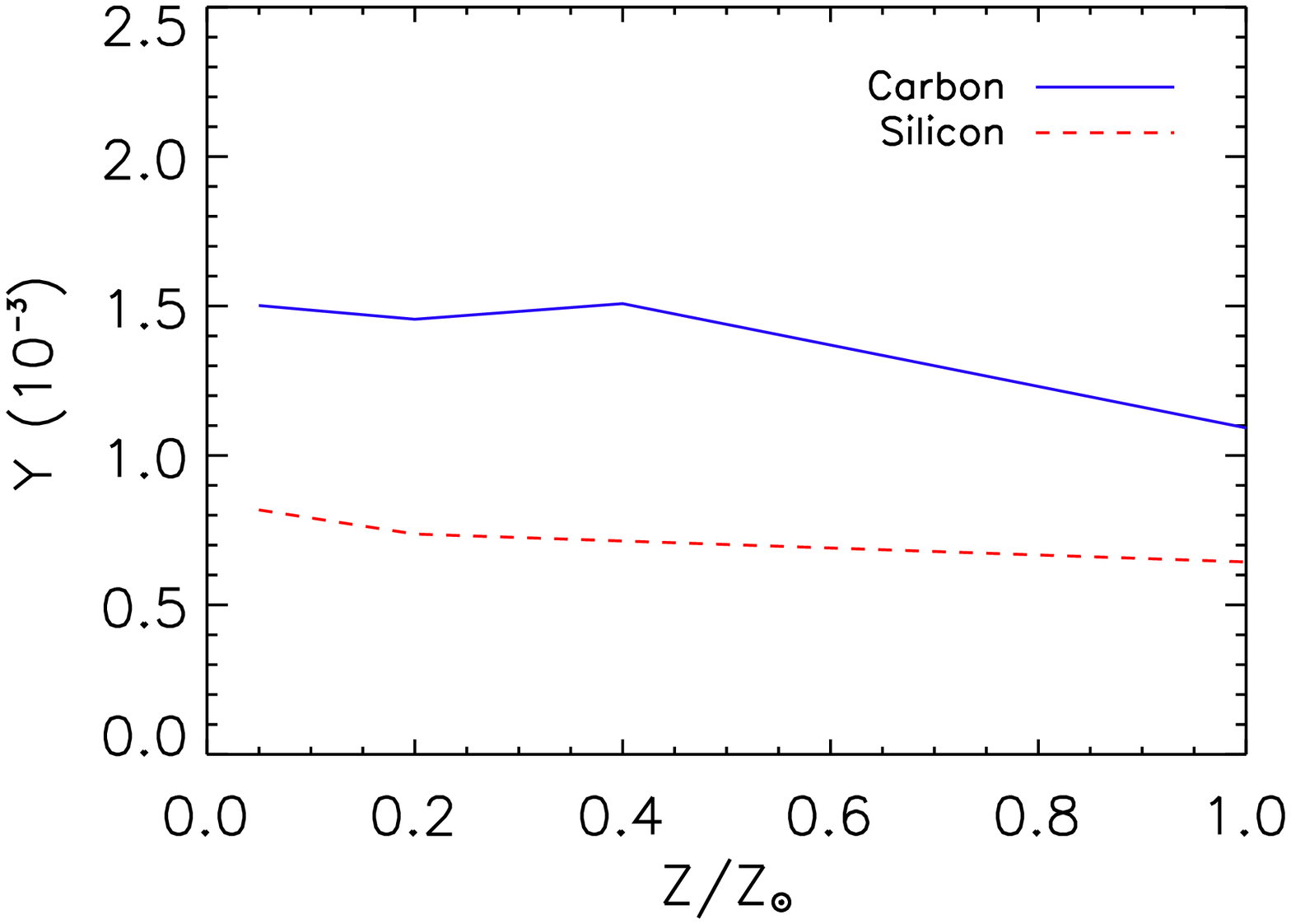}
\caption{Upper panel: Dust condensation efficiency $f_\mathrm{in,X}$ as a function of metallicity $Z$. Bottom panel: Mass fraction of newly produced metal $\mathcal{Y}_\mathrm{X}$ as a function of metallicity $Z$. In both panels, solid and dashed lines indicate carbon and silicon, respectively.}
\label{Fig:yield_z}
\end{center}
\end{figure}

To model the evolution of the two species (silicate and carbonaceous dust) separately, we also need to estimate the fraction of newly produced metals ($\mathcal{Y}_\mathrm{X}$) and the dust condensation efficiency ($f_\mathrm{in,X}$) for C and Si, by adopting the same yield data as in \citet{Asano:2013aa}.
We adopt the metal and dust yields of AGB stars for a range of progenitor mass at the zero-age main sequence, $m=1$--8 M$_{\solar}$ from \citet{Karakas:2010aa} and \citet{Zhukovska:2008aa}, respectively, and the metal and dust yields of SNe for $m=8$--40 M$_{\solar}$ from \citet{Kobayashi:2006aa} and \citet[][with hydrogen number density 1 cm$^{-3}$ and the unmixed helium core]{Nozawa:2007aa}.
For a given metallicity, $Z$,
$\mathcal{Y}_\mathrm{X}$ and $f_\mathrm{in,X}$ are evaluated as
\begin{equation}
  \mathcal{Y}_\mathrm{X} = \int _{m_t}^{ 40~\mathrm{M}_{\odot} }{ m_{Z,\mathrm{X}}(m)\phi (m)\,\mathrm{d}m },
  \label{eq:yeild}
\end{equation}
and 
\begin{equation}
  f_\mathrm{in, X} =\frac{\int _{m_t}^{ 40~\mathrm{M}_{\odot} }{ m_\mathrm{d,X}(m)\phi (m)\,\mathrm{d}m } }  { \int _{m_t}^{ 40~\mathrm{M}_{\odot} }{ m_{Z,\mathrm{X}}(m)\phi (m)\,\mathrm{d}m }},
  \label{eq:f_in}
\end{equation}
where $m_t$ is the turn-off mass at galaxy age $t$, $\phi (m)$ is 
the stellar initial mass function (IMF), $m_{Z,\mathrm{X}}(m)$ is 
the total mass of newly produced element X as a function of $m$, 
and $m_\mathrm{d,X}$ is the mass of newly produced element X 
condensed into dust as a function of $m$.
Because $\mathcal{Y}_\mathrm{X}$ and $f_\mathrm{in,X}$ are 
much less sensitive to $Z$ than to $m$, 
the values averaged for metallicity between 0.01 $Z_{\solar}$ and 1 $Z_{\solar}$
are used for simplicity (see Fig. \ref{Fig:yield_z}). 
The values of $\mathcal{Y}_\mathrm{X}$ and $f_\mathrm{in,X}$ are listed in 
Table \ref{table:Adopted_quantities}.
We adopt a Salpeter IMF [$\phi (m)\propto m^{-2.35}$] with 
a stellar mass range of 0.1--100 M$_\odot$.
We integrate the metal and dust 
yields up to 40 M$_{\solar}$, assuming that
stars heavier than 40 M$_{\solar}$ do not eject any mass into the ISM \citep{Heger:2003aa}.
The returned fraction is estimated as
\begin{equation}
  \mathcal{R} = \int _{m_t}^{ 40~\mathrm{M}_{\odot} }{ [m - w(m)]\phi (m)\,\mathrm{d}m },
  \label{eq:return_f}
\end{equation}
where the remnant mass, $w(m)$, is adopted from the fitting formula 
provided by \citet{Inoue:2011aa}.
We adopt $\mathcal{R} = 0.25$, the same value as
in our previous work (H15; see also Appendix A in \citealt{Hirashita:2011aa}).

\begin{figure}
\begin{center}
\includegraphics[width=0.45\textwidth]{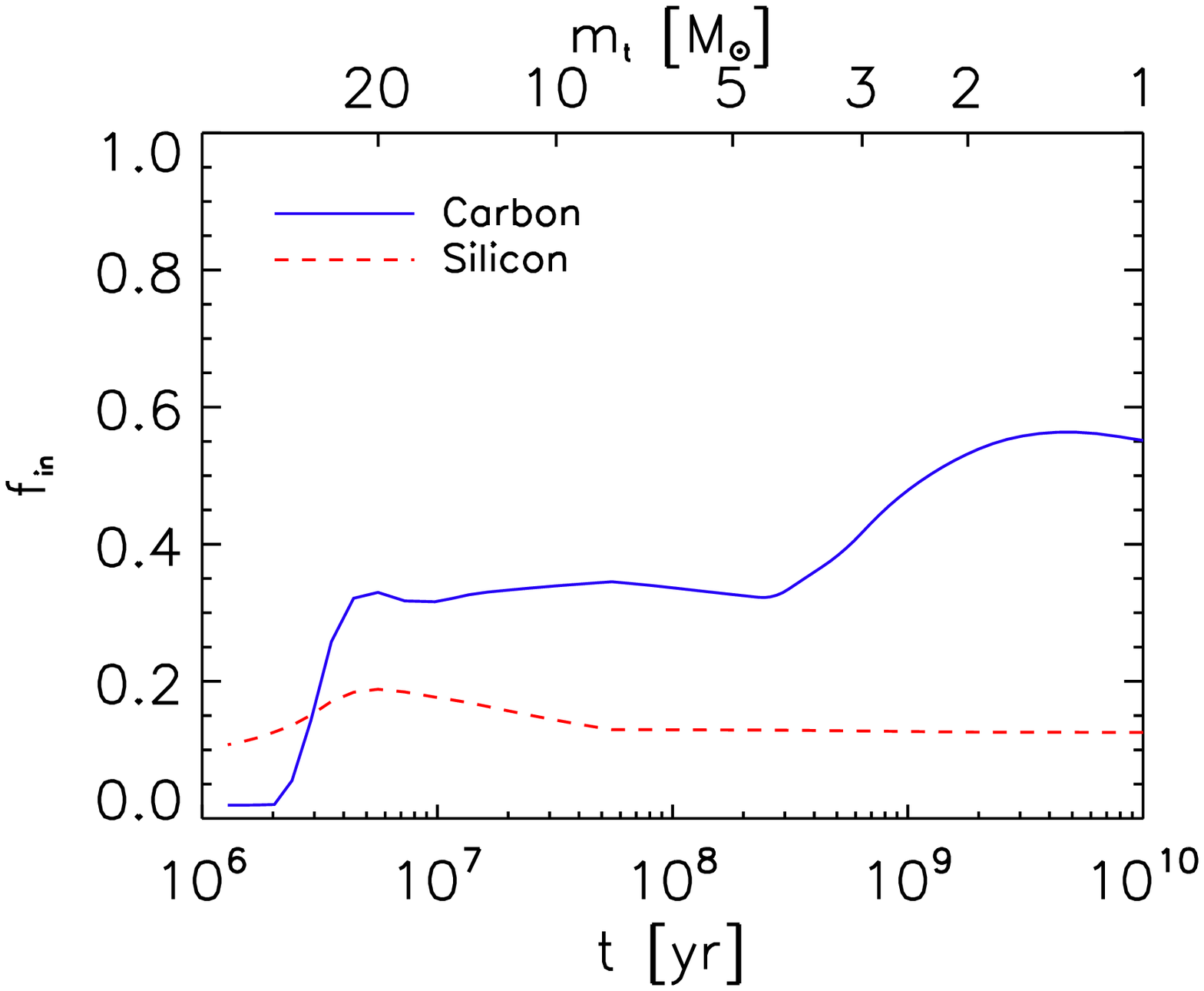}
\includegraphics[width=0.45\textwidth]{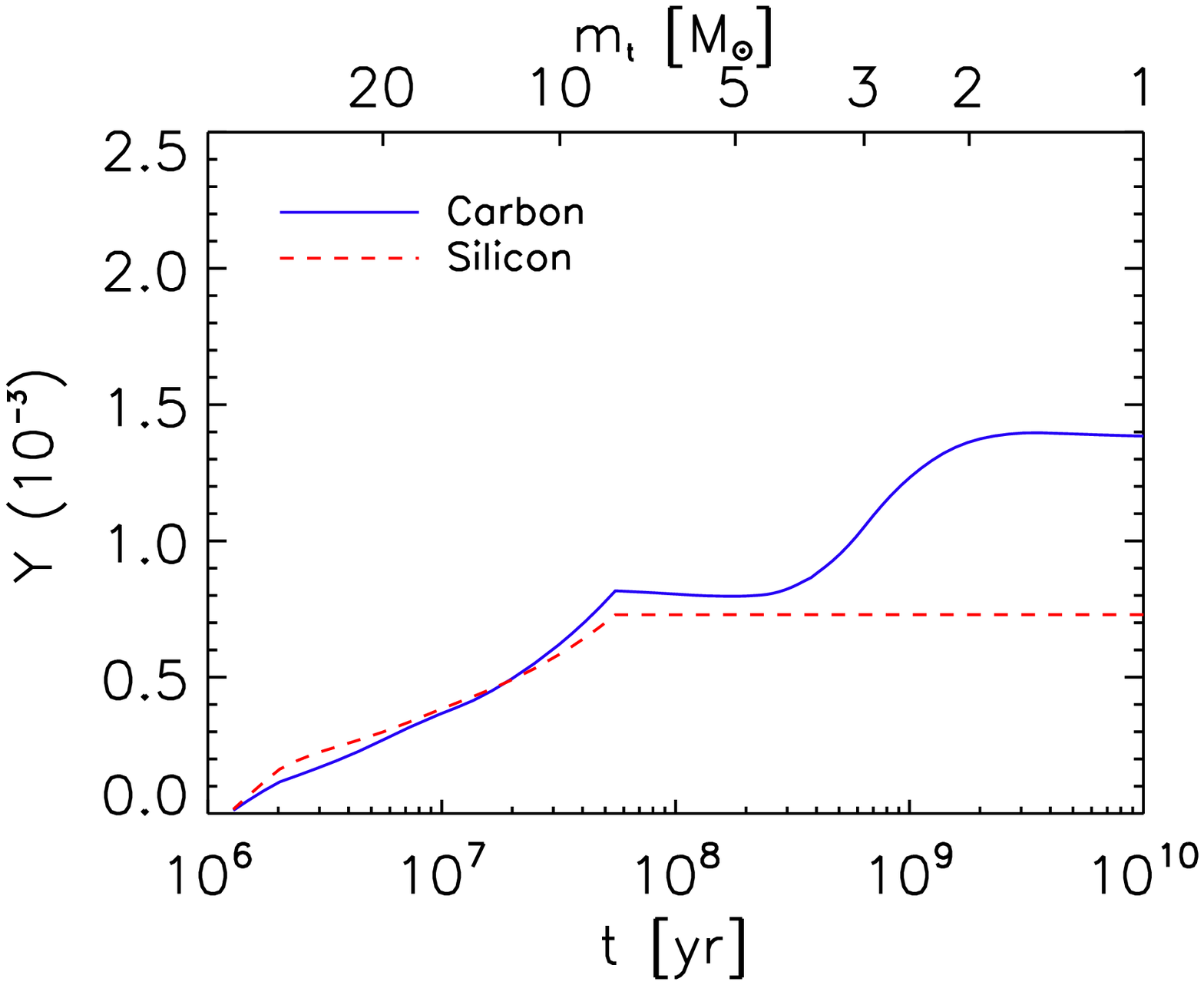}
\caption{
Upper panel: Dependence of the dust condensation efficiency $f_\mathrm{in,X}$ on the age/the turn-off mass. 
Bottom panel: Dependence of the mass fraction of newly produced metals $\mathcal{Y}_\mathrm{X}$ on the age/the turn-off mass. In both panels, solid and dashed lines indicate carbon and silicon, respectively.}
\label{Fig:turnoffmass}
\end{center}
\end{figure}

In Fig. \ref{Fig:turnoffmass}, we show the dependence of 
metallicity-averaged $\mathcal{Y}_\mathrm{X}$ and $f_\mathrm{in,X}$ 
on the age $t$ (or $m_t$).
We observe that
the change of those parameters with the age is
within a factor of 2 as long as
$t\gtrsim 10^9$ yr. This means that, as long as we are interested in the MW, LMC, and
SMC, whose stellar mass has built up over the time comparable to the
cosmic age ($t\gtrsim 10^9$ yr), the results are not sensitive to the choice of $m_t$.
Thus, we assume the constant values listed in Table \ref{table:Adopted_quantities}
for $\mathcal{Y}_\mathrm{X}$ and $f_\mathrm{in,X}$,
where we choose the turn-off mass $m_t =$ 1 M$_{\solar}$.

The dust yield adopted above may still be uncertain.
Alternative dust yield data are available in
\citet{Bianchi:2007aa} for SNe and \citet{Ventura:2012aa}
for AGB stars.
In fact, the shape of
extinction curve is not affected by the detailed choice
of dust yield because of the following reason. If the
dust formation is dominated by stellar sources, the extinction curve
is flat, since we assume that stars produce only large grains.
Thus, the dust composition is of second importance in
determining the shape of extinction curve. If the grain
size distribution is governed by interstellar processing
(i.e.\ processes other than stellar dust formation),
the fraction of metals locked into dust is not determined
by $f_\mathrm{in,X}$ any more but is dominated by
accretion and destruction (\cite{Inoue:2011aa}; H15). 
Therefore, adopting other dust yield data does not
affect our conclusion in this paper.

\begin{table*}
\begin{center}
\begin{minipage}{130mm}
\caption{Adopted quantities.}
\label{table:Adopted_quantities}
\begin{tabular}{lccccccc}
\hline
Species & X & $f_\mathrm{X}$ & $m_\mathrm{X} $ (amu) & $(\mathrm{X/H})_{\odot}$ & $\rho_\mathrm{X}~(\mathrm{g~cm^{-3}})$ & $f_\mathrm{in,X}$ & $\mathcal{Y}_\mathrm{X}$\\
\hline
Carbonaceous dust & C & 1 & 12 & $2.88 \times 10^{-4}$ & 2.24$^a$ & 0.55  & $1.38 \times 10^{-3}$ \\
Silicate & Si & 0.166 & 28.1 &  $4.07 \times 10^{-5}$ & 3.5 & 0.13 & $7.3 \times 10^{-4}$ \\
\hline
\end{tabular}
\\
$^a$Amorphous carbon is introduced for the fitting of the SMC/the LMC 
extinction curves, using $\rho_\mathrm{C} = 1.81~\mathrm{g~cm^{-3}}$
\citep{Zubko:2004aa}.
\end{minipage}
\end{center}
\end{table*}%

\subsection{Extinction curve}\label{subsec:extinc}
The whole range of grain size is represented by only two
populations (large and small grains), while the grain
size distribution itself is necessary in calculating
extinction curves. Thus, in order to calculate extinction curves
in the framework of two-size approximation,
we still
need to assume a specific functional form for the grain
size distribution, as a result of a trade-off between the simplicity of the
two-size approximation and the detailed treatment of grain size distribution.
Following H15, we adopt a modified-lognormal function for the
grain size distribution:
\begin{equation}
  n_{i,\mathrm{X}}(a)=\frac { C_{i,\mathrm{X}} }{ a^{ 4 } } \exp\left\{ - \frac { \left[ \ln(a/a_{ 0,i }) \right] ^{ 2 } }{ 2\sigma^2 }  \right\},
\end{equation}
where subscript $i$ indicates the small ($i = \mathrm{s}$) or large 
($i = \mathrm{l}$) grain component, $C_{i}$ is the normalization 
constant and $a_{0,i}$, and $\sigma$ are the central grain radius 
and the standard deviation, respectively.
We adopt $a_\mathrm{0,s} = 0.005~\micron$, $a_\mathrm{0,l} = 0.1~\micron$ and $\sigma = 0.75$, 
and determine the normalization constants by
\begin{equation}
  {\mu}m_\mathrm{H}\mathcal{D}_{i,\mathrm{X}}=\int_{0}^{\infty}\frac{4}{3}a^3\rho_\mathrm{X}n_{i,\mathrm{X}}(a)\,\mathrm{d}a,
\end{equation}
where $\mu = 1.4$ is the gas mass per hydrogen nucleus, 
$m_\mathrm{H}$ is the mass of hydrogen atom and $\rho_\mathrm{X}$ 
is the material density of dust grains
(given in Table \ref{table:Adopted_quantities}).
H15 confirmed that, if the small-to-large grain abundance ratio is the same
as the MRN grain size distribution, the above functional form correctly
reproduces the MW extinction curve.
\citet{Weingartner:2001aa} made more efforts of precise fitting to the
MW extinction curve. Although the detailed functional form is different,
their size distributions for the mean MW extinction curve still have
mass ratios of large grains to small grains which are similar to the MRN size distribution.
It is worthy pointing out that the peaks of the two lognormal components are located
at similar grain sizes to the two peaks in the graphite grain size distribution
of \citet{Weingartner:2001aa}.

The extinction at wavelength $\lambda$ in units of magnitude ($A_{\lambda,\mathrm{X}}$) normalized to the column density of hydrogen nuclei ($N_\mathrm{H}$) is written as
\begin{equation}
  \frac{A_{\lambda,\mathrm{X}}}{N_\mathrm{H}}=2.5 {\log} \mathrm{e} \sum_{i}\int_{0}^{\infty}n_{i,\mathrm{X}}(a){\pi}a^2Q_\mathrm{ext}(a,\lambda,\mathrm{X}),
\end{equation}
where $Q_\mathrm{ext}(a,\lambda,\mathrm{X})$ is the extinction coefficient (extinction cross section normalized to the geometric cross section) as a function of grain size, wavelength and species. The total extinction $A_\lambda$ is calculated by summing $A_{\lambda ,\mathrm{X}}$ for all the species.
We calculate $Q_\mathrm{ext}(a,\lambda,\mathrm{X})$ using the Mie theory \citep{Bohren:1983aa} based on the same optical constants for silicate and graphite in \citet{Weingartner:2001aa}. 
All the grains are assumed to be spherical with uniform composition.
For the SMC and LMC, we introduce amorphous carbon (AC) for 
carbonaceous dust as a representative carbonaceous species 
without the 2175 \AA\ feature, and the extinction coefficient 
of amorphous carbon is calculated with the optical constants 
provided by \citet{Zubko:1996aa} (their `ACAR' is adopted).

\subsection{Observational data for extinction curves}

In order to judge whether or not the calculated extinction curves
successfully explain the observed extinction curves,
we compare the extinction curve shape,
which is defined as $A_\lambda /A_V$ (the wavelength at the $V$ band
is 0.55 $\micron$). We introduce $\Delta^2$ to measure the
`distance' between the calculated and observed extinction curves:
\begin{equation}
  \Delta^{ 2 }=\sum_{ i }[(A_{\lambda_i} /A_V)_\mathrm{model}-
(A_{\lambda_i}/A_V)_\mathrm{obs})]^{ 2 },
\label{eq:delta}
\end{equation}
where the subscripts `model' and `obs' indicate the model
and the observational values,
respectively, and the sum is taken for all the sampled wavelengths
$\lambda_i$.

The observed extinction curves are taken from \citet{Pei:1992aa}.
The extinction curves are given at 30 wavelengths from
ultraviolet (UV) to near IR. These wavelengths are used for
$\lambda_i$ above.
\citet{Fitzpatrick:2007aa} presented extinction curves
along various line of sights in the MW. Since these extinction curves
show a dispersion as quantified by \citet{Nozawa:2013aa},
we will later use the dispersion to define a `reasonable' range for the
above distance $\Delta^{2}$.

We surveyed the appropriate parameter ranges listed in
Table \ref{table:parameters}, which follows the values suggested in H15.
H15 suggested these ranges based on previous theoretical dust evolution
studies and confirmed that they reproduce the relation between dust-to-gas
ratio and metallicity in nearby galaxies.
Each model with a certain set of parameters produces the
dust-to-gas ratio $\mathcal{D}_{i,\mathrm{X}}$ 
as a function of metallicity $Z$. We calculate the extinction curve
using the calculated $\mathcal{D}_{i,\mathrm{X}}$ at
the metallicity appropriate for each galaxy.
In this paper, we assume the metallicities of the MW, LMC and SMC 
to be 1 $Z_{\solar}$, 0.5 $Z_{\solar}$ and 0.2 $Z_{\solar}$, respectively
\citep{Russell:1992aa,Korn:2000aa}.

\begin{table}
\caption{Parameter ranges surveyed.}
\begin{center}
\begin{tabular}{lcccc}
\hline 
Process & Parameter & Minimum & Middle & Maximum \\
\hline
Star formation & $\tau_\mathrm{SF}$ & $5\times10^{8}$ yr$^*$  &  $5\times10^{9}$ yr  &  $5\times10^{10}$ yr   \\
Shattering  & $\tau_\mathrm{sh,0}$ & $10^{7}$ yr$^*$  &  $10^{8}$ yr  &  $10^{9}$ yr   \\
Coagulation & $\tau_\mathrm{co,0}$ & $10^{6}$ yr$^*$  &  $10^{7}$ yr  &  $10^{8}$ yr   \\
Accretion & $\tau_\mathrm{cl}$ & $10^{6}$ yr  &  $10^{7}$ yr  &  $10^{8}$ yr$^*$   \\
SN destruction & $\beta_\mathrm{SN}$  &  4.82 &  9.65$^*$ &  19.3 \\
\hline
\end{tabular}
\end{center}
$^*$Fiducial case (see Section \ref{subsec:param}).
\label{table:parameters}
\end{table}%

\section{results}

\subsection{Dust-to-gas ratio}\label{subsec:result_D}
\begin{figure}
\begin{center}
\includegraphics[width=0.45\textwidth]{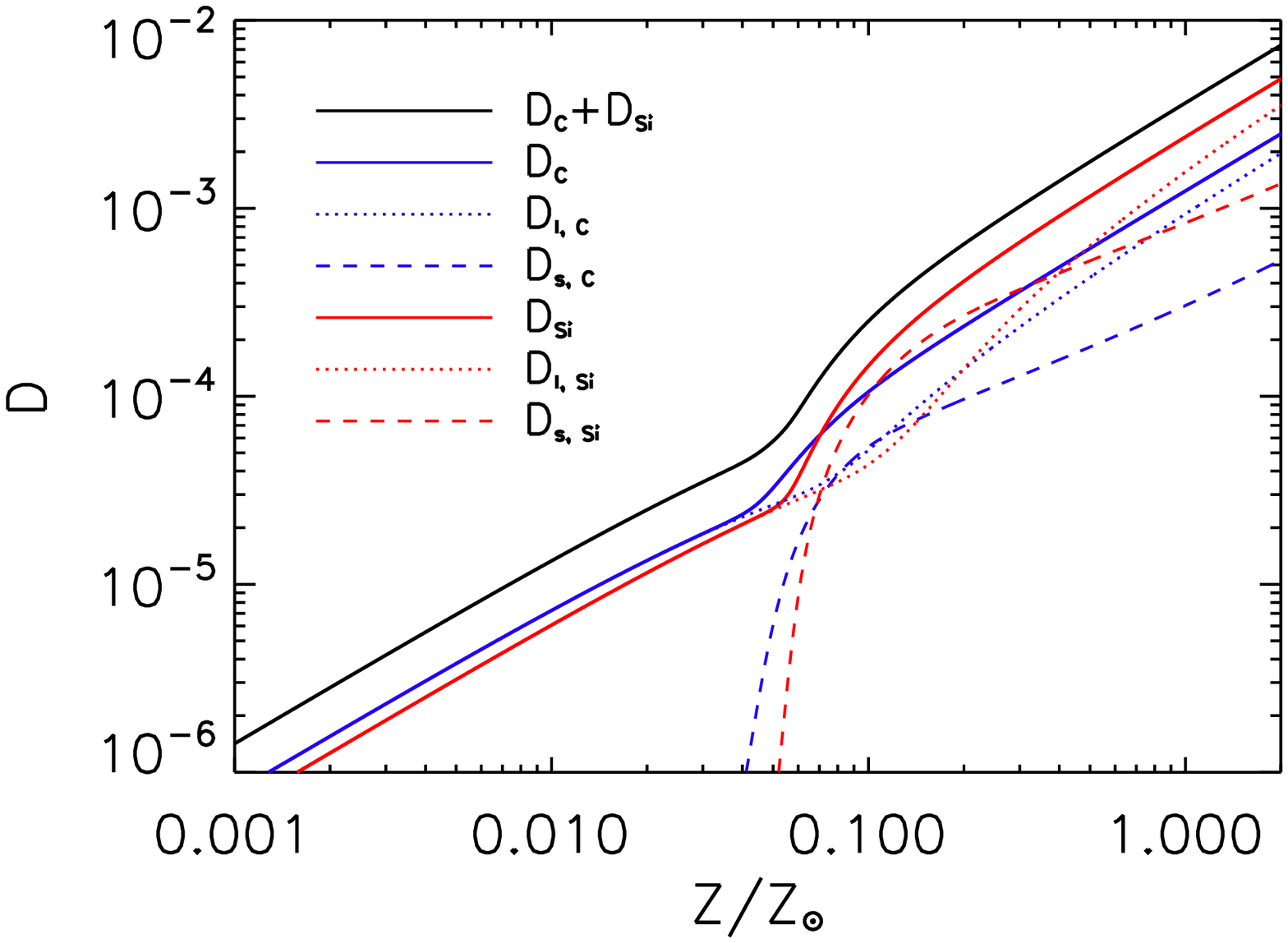}
\includegraphics[width=0.45\textwidth]{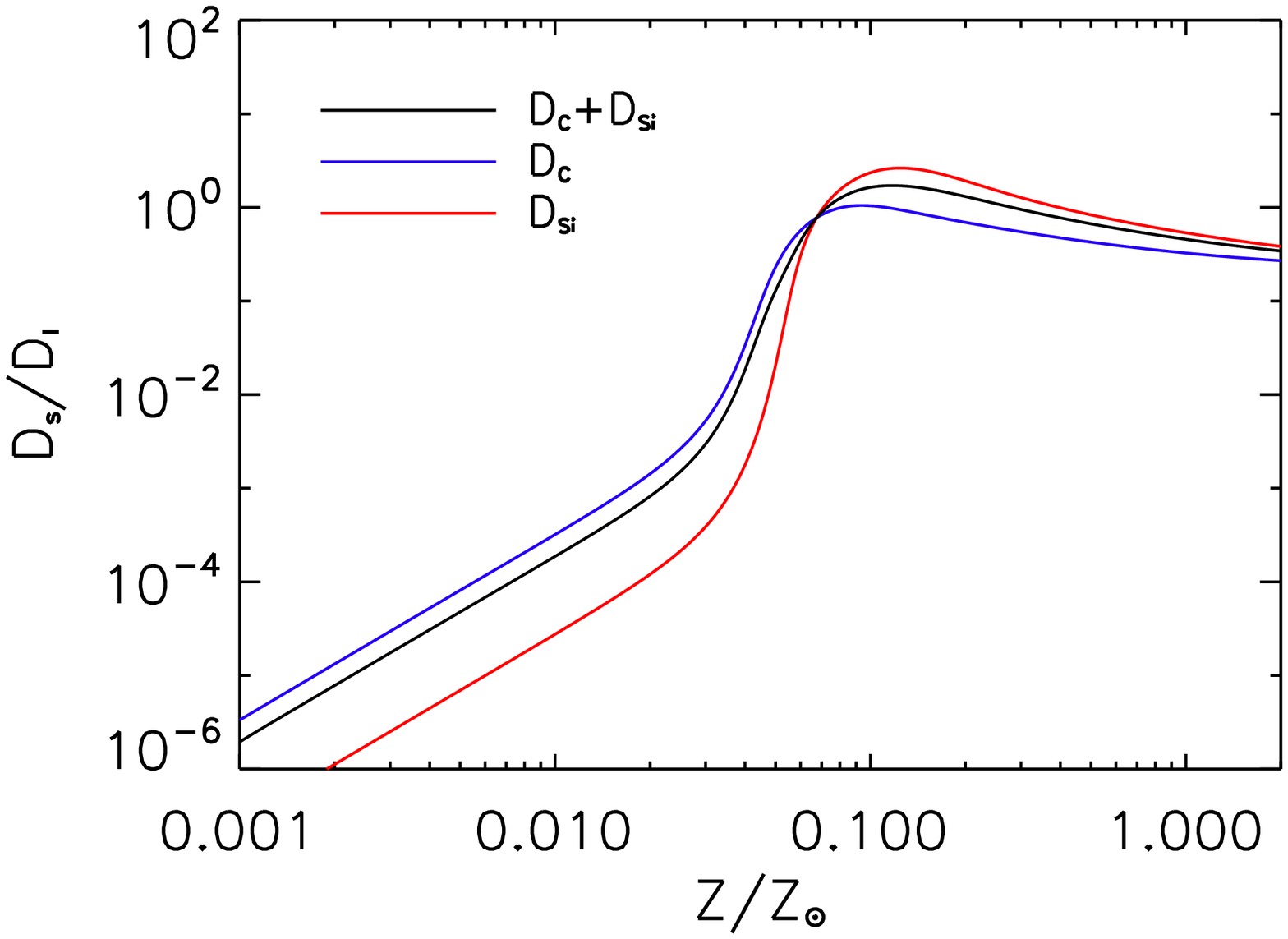}
\caption{Upper panel: Dust-to-gas mass ratio as a function of metallicity $Z$ normalized to the solar metallicity. We show the dust-to-gas ratios of carbonaceous dust (blue lines) and silicate (red lines) separately and the small grain dust-to-gas ratio (dashed lines) and large grain dust-to-gas ratio (dotted lines). The black solid line shows the total dust-to-gas ratio. Bottom panel: Small-to-large grain abundance ratio as a function of metallicity $Z$.
The colors are the same as in the upper panel. }
\label{Fig:Dust-to-gas_ratio}
\end{center}
\end{figure}

The parameter dependence of the evolution of dust-to-gas ratio
has already been investigated in H15.
We only show an example of the evolution of dust-to-gas ratio
for the fiducial parameter values given in Table \ref{table:parameters}
(see Section \ref{subsec:param} for the choice of the fiducial parameter
set).
In Fig.\ \ref{Fig:Dust-to-gas_ratio}, we show $\mathcal{D}_\mathrm{l,X}$, $\mathcal{D}_\mathrm{s,X}$, $\mathcal{D}_\mathrm{X}$($\equiv\mathcal{D}_\mathrm{l,X}+\mathcal{D}_\mathrm{s,X}$) and $\mathcal{D}$($\equiv\mathcal{D}_\mathrm{C}+\mathcal{D}_\mathrm{Si} $) as a function of $Z$.
At low metallicity, the dust production is dominated by the stellar sources
(i.e.\ SNe and AGB stars), which are assumed to form large grains.
At this stage, $\mathcal{D_\mathrm{X}}\simeq f_{\mathrm{in,X}}Z$ until SN destruction is 
strong enough to suppress the increasing rate.
This suppression by destruction is seen around $Z\sim 0.03 Z_{\solar}$
in Fig.\ \ref{Fig:Dust-to-gas_ratio}.
In the meantime, shattering continuously transforms large 
grains into small grains as seen in the initial rise of
the small-to-large grain abundance ratio ($\mathcal{D}_\mathrm{s,X}/\mathcal{D}_\mathrm{l,X}$).
Shattering is thus important for the first production of
small grains. Subsequently,
accretion efficiency increases the abundance of small grains
as a consequence of the enhancement of grain surface area by
shattering as well as the increase of metallicity
\citep[i.e. abundance of accreting materials; see also][]{Dwek:1980aa,Inoue:2011aa,Kuo:2012aa,Mattsson:2014aa}.
The silicate abundance becomes greater than carbonaceous dust 
abundance at this epoch. The dust mass increase by accretion
saturates after a significant fraction of gas-phase metals
are consumed.
Coagulation becomes active with the enhancement of small grain
abundance; as a result of coagulation, the small-to-large grain
abundance ratio decreases.
We refer the interested reader to H15 for the detailed discussions
on the evolution of dust-to-gas ratio and small-to-large
grain abundance ratio.

\subsection{Parameter dependence of extinction curve}\label{subsec:param}

\begin{figure*}
\begin{center}
\includegraphics[width=\textwidth]{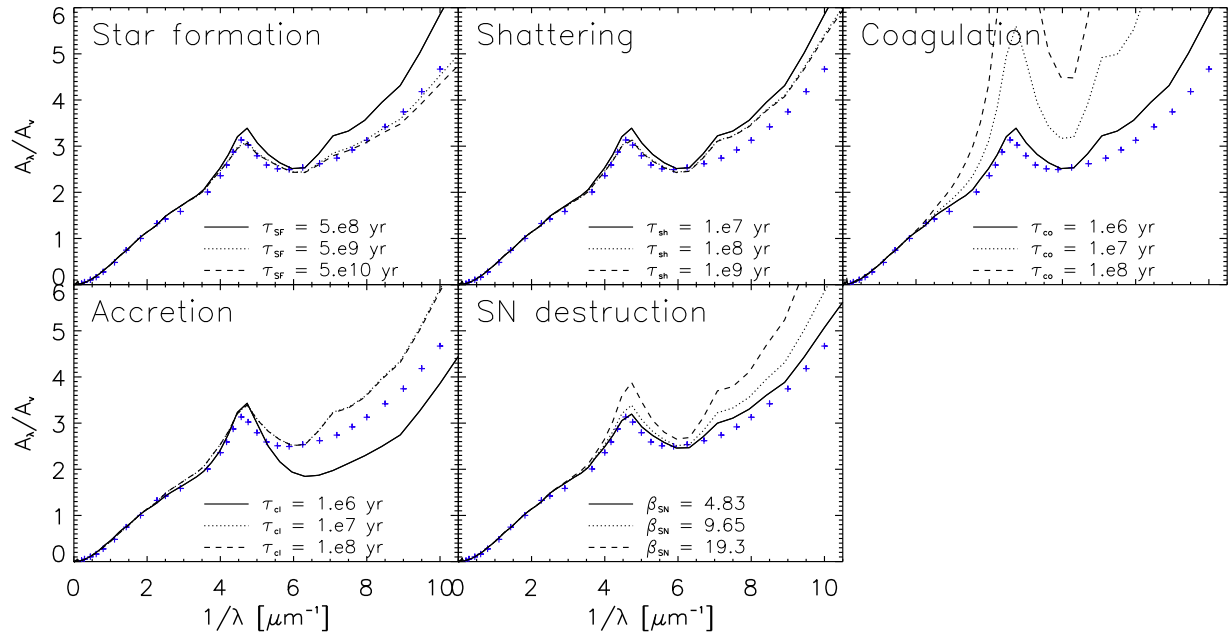}
\caption{Effect of each parameter on extinction curves. 
For the parameters other than the one specified in each panel, we adopt the fiducial values listed in Table \ref{table:parameters}. Cross symbols show the observed mean extinction curve of the MW taken from \citet{Pei:1992aa}. Solid, dotted and dashed lines correspond to the value of the parameter listed in the legend in each panel.}
\label{Fig:MW_each_parameter}
\end{center}
\end{figure*}

For the MW extinction curve, we use the results for each set of parameters at metallicity $Z = Z_{\solar}$, which is appropriate for the MW, and calculate
the extinction curve by adopting graphite and silicate for the dust species.
The calculated MW extinction curves are compared with the observed
curve \citep{Pei:1992aa}.
We examine the parameter dependence of extinction curve in Fig.\ \ref{Fig:MW_each_parameter}.
To this aim, we first choose a set of parameters that roughly fits the
observed MW extinction curve.
This parameter set is referred to as the fiducial case (Table \ref{table:parameters}).
Note that, as shown later, there are other sets of parameters that
give satisfactory fitting to the MW extinction curve; thus, the
aim of taking the fiducial parameters is only to clarify the dependence
on the parameters.
Fig. \ref{Fig:MW_each_parameter} shows the extinction curve for the fiducial parameter set along with the variation caused by the change of each individual parameter.

We observe that the extinction curves are sensitive to the star formation, coagulation and accretion time-scales.
Although Fig.\ \ref{Fig:MW_each_parameter} shows just example cases,
it represents a general trend of extinction curves with
varying parameters. We focus on the 2175 \AA\ bump and the UV slope 
as prominent signatures of extinction curves. 
In general, shorter star-formation and shattering time-scales ($\tau_\mathrm{SF}$ 
and $\tau_\mathrm{sh}$), a longer coagulation time-scale ($\tau_\mathrm{co}$) 
and a longer cloud lifetime ($\tau_\mathrm{cl}$) produce the extinction 
curves with a more prominent bump and a steeper slope because the
small-to-large grain abundance ratio is larger.
Shorter $\tau_\mathrm{sh}$ (high shattering efficiency)
produces more small grains. Coagulation has an opposite effect:
short $\tau_\mathrm{co}$ means efficient conversion of
small grains into large gains, which smears out the 
signatures of extinction curve. The abundance of small grains
increases more efficiently in a 
condition of longer $\tau_\mathrm{cl}$, which makes a stronger bump and
a steeper slope. 
Shorter $\tau_\mathrm{SF}$ decreases both shattering and coagulation 
efficiencies (note that $\beta_\mathrm{sh}$ and $\beta_\mathrm{co}$
are both proportional to $\tau_\mathrm{SF}$;
Section \ref{subsec:twosize}), but the effect of shattering appears
more prominently since coagulation takes place only after shattering
(Section \ref{subsec:result_D}). The SN destruction efficiency
$\beta_\mathrm{SN}$ shows a minor effect compared to other parameters;
however, $\alpha_\mathrm{X}$, enhancement factor of SN destruction for 
small grains, proves to be important when we model the SMC/LMC extinction 
curves later.

\subsection{The MW}

We examine the $3^5=243$ combinations of parameter values listed in
Table \ref{table:parameters}. Among them, we choose the cases with
small $\Delta^2$  (equation \ref{eq:delta}) for satisfactory fits.
We adopt $\Delta^2\leq 8$ for the criterion of
good fit. As shown below, if we choose the extinction curves
satisfying this criterion, the calculated extinction curves are
roughly within the observed dispersion of the extinction curves
in the MW in various lines of sight. This criterion also
empirically works for the LMC and SMC as shown below. Since this criterion is only
empirically imposed, we also show all the calculated extinction curves
satisfying the criterion to visually confirm that
the extinction curves calculated are actually near to the
observed one.

\begin{figure}
\begin{center}
\includegraphics[width=0.45\textwidth]{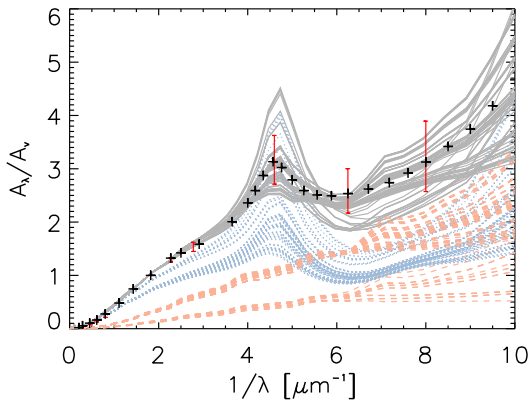}
\caption{Simulated MW extinction curves satisfying $\Delta^{2} \leq 8$. Cross symbols are the same as in Fig. \ref{Fig:MW_each_parameter} and vertical bars are the 1$\sigma$ dispersion taken from \citet{Fitzpatrick:2007aa}. Gray lines present the total extinction curves, and blue dotted and orange dashed lines are the graphite and silicate component of extinction curves.
}
\label{Fig:MW_good_set}
\end{center}
\end{figure}

In Fig.\ \ref{Fig:MW_good_set}, we plot all the model extinction curves
satisfying the criterion $\Delta^2\leq 8$. We also show the dispersions
of the MW extinction curves in various lines of sight
\citep{Fitzpatrick:2007aa, Nozawa:2013aa}. We observe that the
extinction curves selected with $\Delta^{2} \leq 8$ roughly have a
comparable dispersion to the observed ones. In other words,
$\Delta^{2} \leq 8$ can be regarded as an acceptable range for
the MW condition.
We overproduce the dispersion for the 2175 \AA\ bump feature,
but this could be improved by a slight inclusion of AC, which
does not show the 2175 \AA\ bump feature.

In total, 35 per cent of the parameter sets satisfy
the criterion, which implies that our model successfully contains
the processes and parameter ranges that are suitable for the MW
condition.
On the other hand, the fact that a large number of
parameter sets can reproduce the MW extinction curve indicates
degeneracy; that is, the dust evolution
history that reproduces the MW extinction curve is not unique.
Nevertheless, we will show later that there is some preferred
parameter space, and we will discuss the characteristics of the parameter sets
that satisfy the criterion in Section \ref{subsec:constraint}.

\subsection{The SMC}\label{subsec:smc}
The SMC extinction curve is very different from the MW curve in that it does not show a prominent 2175 \AA\ bump, and the LMC extinction curve is intermediate between these two curves  \citep[e.g.][]{Fitzpatrick:1985aa}. Therefore, we examine the SMC extinction curve first, and model the LMC curve as an intermediate case later.
There is one thing we should keep in mind.
The 2175 \AA\ bump is absent in the SMC bar region, but a weak bump exists in the SMC wing region
(Gordon et al.\ 2003),
which implies a spatial variation of carbonaceous material properties
among graphite, AC, and polycyclic aromatic hydrocarbons (PAHs)
(Li \& Draine 2002).
This variation may be driven by UV radiation (Jones 2009), but the mechanism of
UV processing is not fully understood yet. In this paper, we do not include
such processing explicitly, but simply focus on how the
bumpless extinction curve is reproduced by a certain carbonaceous material.

\begin{figure}
\begin{center}
\includegraphics[width=0.45\textwidth]{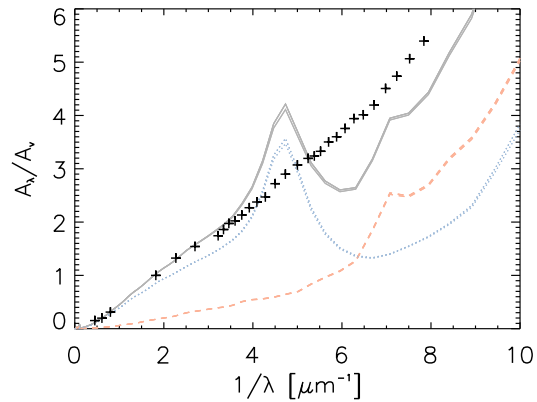}
\caption{Five best fitting extinction curves to the SMC with the same
parameter sets and dust properties adopted for the MW. The crosses are the observed the SMC extinction curve taken from \citet{Pei:1992aa}. All the models have $\Delta^2>8$, and thus the fit is not satisfactory. Lines and colors are the same as Fig. \ref{Fig:MW_good_set}. Extinction curves are overlapping because of the degeneracy of parameters. }
\label{Fig:SMC_MWpara_set}
\end{center}
\end{figure}

We surveyed all the parameter sets and examined the goodness of fit for the SMC extinction curve at $Z=0.2 Z_{\solar}$ with the same dust species (silicate and graphite) as in the MW. However, we did not find any curve that satisfies $\Delta^2\leq 8$. In Fig.\ \ref{Fig:SMC_MWpara_set}, we show the five best fit models to the SMC extinction curve. The averaged $\Delta^2$ for the best five fitting results (denoted as $\overline{\Delta^2}$) is $\overline{\Delta^2}= 15.8$.

As shown in Fig.\ \ref{Fig:SMC_MWpara_set},
the largest discrepancy between the models and observation appears at UV wavelengths. The observed SMC extinction curve has no prominent 2175 \AA\ bump feature and has a steep slope. In our models, the 2175 \AA\ feature is due to small graphite grains and the steep UV slope is mostly due to small silicate grains. To reproduce the SMC extinction curves, thus, we propose two possible modifications: one is to use AC instead of graphite and the other is to
increase the SN destruction efficiency for small carbonaceous grains ($\alpha_\mathrm{X}$ in equation \ref{eq:small} with X = C).
For the second solution,
the 2175 \AA\ bump feature should be suppressed while the steepness of the UV extinction curve
should be kept, which indicates less small carbonaceous dust and enough small silicates; thus, we set $\alpha_\mathrm{C} = 0.1$ to destroy small carbonaceous dust grains efficiently with $\alpha_\mathrm{Si} = 1$ remaining the same to preserve small silicate grains.
In the following, we examine these two possibilities.

\subsubsection{AC}
\begin{figure}
\begin{center}
\includegraphics[width=0.45\textwidth]{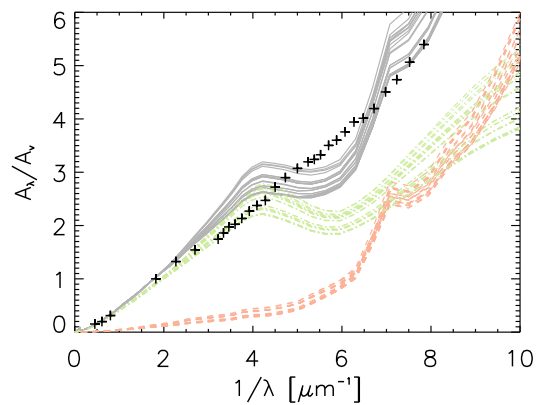}
\caption{Simulated SMC extinction curves satisfying $\Delta^2\leq 8$ with adopting AC for carbonaceous dust. $\overline{\Delta^2} = 5.7$ and 10\% parameter sets satisfy $\Delta^2\leq 8$. The observational data points are the same as in Fig.\ \ref{Fig:SMC_MWpara_set}.
Gray lines present the total extinction curves, green dash-dotted and orange dashed lines are the AC and silicate component of extinction curves.}
\label{Fig:the SMC_AC_set}
\end{center}
\end{figure}

To solve the problem of the 2175 \AA\ bump being too prominent, we introduce amorphous carbon \citep{Zubko:1996aa}, a different type of carbonaceous dust, which has no 2175 \AA\ bump feature. The same material is also used by \citet{Nozawa:2015aa} to explain the bumpless extinction curve of a high-redshift quasar.
10 per cent (25 parameter sets) of the 243 parameter sets satisfy $\Delta^2\leq 8$.
All these extinction curves with $\Delta^2\leq 8$ are shown in
Fig.\ \ref{Fig:the SMC_AC_set}. The extinction curves calculated with AC are closer to the observed SMC extinction curve than with graphite.
However, AC still make a feature at wavelengths between 1,500 and 2,500 \AA.
The averaged $\Delta^2$ for the five best fit results is $\overline{\Delta^2} = 5.7$, much smaller than the above (15.8). Therefore, using AC for carbonaceous dust improves the fit to the SMC extinction curve.

\subsubsection{$\alpha_\mathrm{C} = 0.1$ with graphite}
\begin{figure}
\begin{center}
\includegraphics[width=0.45\textwidth]{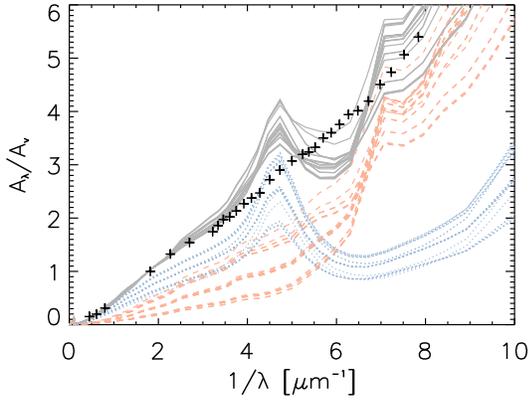}
\caption{Same as Fig.\ \ref{Fig:the SMC_AC_set} but with $\alpha_\mathrm{C}=0.1$ and graphite for the carbonaceous species. $\overline{\Delta^2} = 4.0$ and 9\% parameter sets satisfy $\Delta^2\leq 8$. The observational data points are the same as in Fig.\ \ref{Fig:SMC_MWpara_set}.
Gray lines present the total extinction curves, and blue dotted and orange dashed lines are the graphite and silicate component of extinction curves.}
\label{Fig:the SMC_SN_set}
\end{center}
\end{figure}

We propose a higher SN destruction efficiency for small carbonaceous dust grains than small silicates by introducing a small value of $\alpha_\mathrm{C}$
to relatively suppress the abundance of small carbonaceous grains.
This is motivated by the fact that the prominent 2175 \AA\ bump is caused by
small graphite grains: thus, we expect that the SMC extinction curve is
better explained by the stronger destruction of small carbonaceous grains.
Recall that we adopted $\alpha_\mathrm{X}=1$ for both carbonaceous dust (X = C) and silicate (X = Si) above. Here we give  $\alpha_\mathrm{C} = 0.1$ (an order of magnitude higher destruction rate for small carbonaceous dust) with $\alpha_\mathrm{Si}=1$ unchanged. Fig.\ \ref{Fig:the SMC_SN_set} shows the extinction curves satisfying $\Delta^2\leq 8$. We observe that the fitting is improved compared with the case of $\alpha_\mathrm{C}=1$, although the 2175 \AA\ bump is still prominent. The bump remains prominent because accretion still contributes to
enhancing the small grain abundance.
We find 9 per cent (23 parameter sets) out of the 243 models satisfying $\Delta^2\leq 8$.
The average $\Delta^2$ for the 5 best fit cases is $\overline{\Delta^2} = 4.0$, 
which is smaller than the above case; however, the persistence of bump 
indicates that enhancing the destruction of the bump carrier 
(small graphite grains in our case) is not a probable way of 
explaining the SMC extinction curve.

\subsubsection{$\alpha_\mathrm{C} = 0.1$ with AC}
\begin{figure}
\begin{center}
\includegraphics[width=0.45\textwidth]{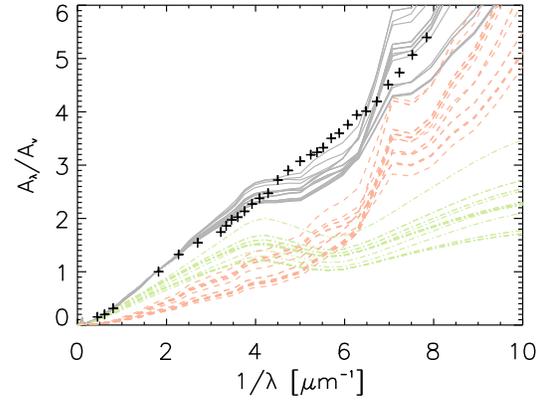}
\caption{Same as Fig.\ \ref{Fig:the SMC_AC_set} but with $\alpha_\mathrm{C}=0.1$ and AC. We show not only the five best fits but also all the calculated curves satisfying $\Delta^2 \leq 8$. $\overline{\Delta^2} = 2.6$ and 11\% parameter sets satisfy $\Delta^2\leq 8$. 
The observational data points are the same as in Fig.\ \ref{Fig:SMC_MWpara_set}.
Gray lines present the total extinction curves, green dash-dotted and orange dashed lines are the AC and silicate component of extinction curves.}
\label{Fig:the SMC_good_set}
\end{center}
\end{figure}

Adopting AC and the small $\alpha_\mathrm{C}(=0.1)$ at the same time, 
we calculate the extinction curve. We show the cases with satisfactory 
fit ($\Delta^2\leq 8$) in Fig.\ \ref{Fig:the SMC_good_set}.
The average $\Delta^2$ for the five best fitting cases is $\overline{\Delta^2} = 4.0$.
Out of all the possible combinations of parameter values in Table \ref{table:parameters}, we find that 8 per cent, 20 parameter sets, of them satisfy $\Delta^2\leq 8$. Although the percentage is not as high as the MW case, it is worth emphasizing that we find some satisfactory dust evolution models that fit the SMC extinction curve with a common framework of dust enrichment. We discuss the properties of parameter sets giving the good fits in Section \ref{subsec:constraint}.

\begin{table}
\caption{Fitting to the SMC extinction curve with different model.}
\begin{center}
\begin{tabular}{lcc}
\hline
Model & $\overline{\Delta^2}$$^*$  & $\Delta^2\leq 8^{**}$ \\
\hline
 $\alpha_\mathrm{C} = 1$ with graphite & 15.8 & 0 \%  \\
 $\alpha_\mathrm{C} = 1$ with AC & 5.7 & 10 \%  \\
 $\alpha_\mathrm{C} = 0.1$ with graphite & 4.0 & 9 \% \\
 $\alpha_\mathrm{C} = 0.1$ with AC & 4.0 & 8 \% \\
\hline
\multicolumn{3}{l}{$^*$Average $\Delta^{2}$ of 5 best fitting. } \\
\multicolumn{3}{l}{$^{**}$Fraction of parameter sets satisfying $\Delta^2\leq 8$.}
\end{tabular}
\end{center}
\label{table:the SMC_good_set}
\end{table}%

\subsection{The LMC}\label{subsec:lmc}

\begin{figure}
\begin{center}
\includegraphics[width=0.45\textwidth]{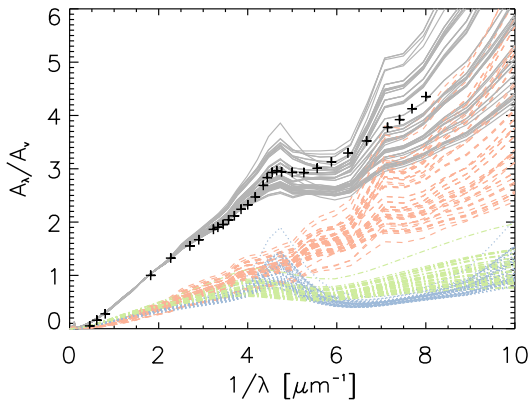}
\caption{Fit to the LMC extinction curve with $\alpha_\mathrm{C}=0.2$ and a mixture of 50 per cent graphite and 50 per cent AC for carbonaceous dust. The calculated curves satisfying $\Delta^2\leq 8$ are shown. Cross symbols show the observed mean extinction curve of the LMC taken from \citet{Pei:1992aa}. Gray lines present the total extinction curves, and blue dotted, green dash-dotted and orange dashed lines are the graphite, AC and silicate component of extinction curves.}
\label{Fig:the LMC_good_set}
\end{center}
\end{figure}

Like the SMC case, 
the model with $\alpha_\mathrm{C}=1$ and using graphite for carbonaceous dust
does not reproduce the LMC extinction curve at $Z=0.5 Z_{\solar}$.
Because its 2175\AA\ bump strength and UV slope are intermediate between
the MW and SMC extinction curves, we treat the LMC as an intermediate case between the MW and SMC. We use the result at $Z = 0.5 Z_{\solar}$, an appropriate ISM metallicity for the LMC \citep{Russell:1992aa}. As intermediate values, we apply $\alpha_\mathrm{C}=0.2$ and a mixture of 50 per cent graphite and 50 per cent AC for the carbonaceous component with the silicate properties unchanged.
We calculate the extinction curves for each parameter set, and show
the extinction curves under the same criterion as the above ($\Delta^2\leq 8$) for the LMC in Fig.\ \ref{Fig:the LMC_good_set}. Among all the possible combinations of parameter values, we find that 27 per cent (66 parameter sets) of them satisfy the criterion. 

The graphite fraction and $\alpha_\mathrm{C}$ used for the fit
of the extinction curves in each galaxy is listed in
Table \ref{table:different_model} with the percentage of
satisfactory fits. We hereafter adopt the graphite fraction and
$\alpha_\mathrm{C}$
in Table \ref{table:different_model} for each galaxy.

\begin{table}
\caption{Tuned Parameters}
\begin{center}
\begin{tabular}{lccc}
\hline
Galaxy & graphite \%$^*$ &  $\alpha_\mathrm{C}$ & $\Delta^2\leq 8^{**}$ \\
\hline
MW &   100 & 1 & 35 \% \\
LMC & 50 &  0.2 & 27 \% \\
SMC & 0 & 0.1 & 8 \% \\
\hline
\end{tabular}
\end{center}
$^*$Mass fraction of graphite for carbonaceous species. The rest is AC. \\
$^{**}$Fraction of parameter sets satisfying $\Delta^2\leq 8$.
\label{table:different_model}
\end{table}%

\section{discussion}

\subsection{Other possible dust species}
Although some detailed dust models incorporating PAHs and 
other complicated dust species have been developed for
extinction curves 
\citep{Li:2001aa,Zubko:2004aa,Draine:2007aa}, 
our conclusions by using only graphite and silicate still hold 
at least qualitatively because of the following reasons. 
Small graphite was classically introduced as being
responsible for the 2175 \AA\ bump
but we can derive the same conclusion as long as the 2175 \AA\
carrier is small grains. Steepness of far UV slope is 
mainly due to small silicate grains.
The requirement of small grains for the UV slope robustly
holds even if we adopt other grain species.
Moreover, by construction of our models, we adopt the
time-scales of various processes in a way independent
of grain species.
Therefore, our
conclusion on the contributions from shattering, coagulation
and accretion, which determine the small-to-large grain abundance
ratio still holds as long as this ratio shapes the extinction curves.

\subsection{Constraint on the parameters}\label{subsec:constraint}

H15 showed that the dust-to-gas ratio produced by our model agrees with the
relation between dust-to-gas ratio and metallicity of nearby galaxies.
In this paper, thus, we have been concentrating extinction curves rather than 
dust abundance.
Since the extinction curves calculated by our models directly reflect the parameters in our dust enrichment models, we here examine whether or not we can get useful constraints on those parameters through the observed extinction curves. In order to show the sensitivity to each parameter, we show the fraction of all the models that satisfy $\Delta^2\leq 8$, which is used as a criterion of good fit.
Because the time-scales of shattering and coagulation are degenerate with the star formation time-scale \citep{Hirashita:2015aa}, we use the $\beta_\mathrm{sh,0}$ ($=\tau_\mathrm{SF}/\tau_\mathrm{sh,0}$) and $\beta_\mathrm{co,0}$ ($= \tau_\mathrm{SF}/\tau_\mathrm{sh,0}$) values to represent the effects of shattering and coagulation. 

In Table \ref{table:good_set}, we show the
percentage of the models satisfying the good-fit criterion with 
the specified parameter fixed and the others changed (over the 
three cases for each parameter) as shown in Table \ref{table:parameters}.
For example, 21 per cent of the models show a satisfactory fit to the Milky Way extinction curve if the star-formation time-scale is fixed to $\tau_\mathrm{SF}=5\times 10^8$ yr.
In general, a higher percentage indicates that the fixed parameter value
is more favored by the observed extinction curve.
Among the processes, coagulation ($\beta_\mathrm{co}$) and accretion ($\tau_\mathrm{cl}$) can be constrained most strongly.
We find that weak coagulation with $\beta_\mathrm{co} \leq  50$ does not reproduce the observed extinction curves for all the MW, LMC and SMC. 
We also observe that strong coagulation with $\beta_\mathrm{co}= 5\times10^4$ find quite a large number of satisfactory fits while such a strong coagulation efficiency is less supported for the SMC. Since coagulation takes place only in the dense ISM \citep{Hirashita:2009aa}, this implies that the dense cloud fraction in the galaxy changes with different metallicities.
Yet, even in the SMC, strong coagulation with
$\beta_\mathrm{co,0}\sim 5\times 10^2$--$5\times 10^3$ is favored.
Necessity of strong coagulation to explain the MW extinction curve is
consistent with more detailed dust evolution models in
\citet{Asano:2014aa} and \citet{Nozawa:2015aa}.

As also observed in Table \ref{table:good_set}, efficient
accretion (larger $\tau_\mathrm{cl}\sim 10^7$--$10^8$ yr) better
explains the observed extinction curves in all the three galaxies. A long cloud lifetime ($\tau_\mathrm{cl} = 10^8$ yr), that is, efficient accretion, is strongly favored for the SMC. The above results indicate that grain growth (accretion and coagulation) is a key process even at the SMC metallicity.
Accretion is the most efficient mechanism of increasing the small
grain abundance (H15), so that it provides the most natural
way of explaining the steepness of the SMC extinction curve.
As shown by \citet{Schneider:2014aa}
in comparison with
the data taken by \citet{Gordon:2014aa},
it is possible that dust produced by stellar sources explains the
total dust budget in the LMC and SMC. However, as also mentioned by
them, if we take dust destruction by SN shocks into account,
an additional source of dust, such as accretion, would be
required (see also \cite{Zhukovska:2013aa}).
The importance of dust growth by accretion for the total dust 
abundance has already been shown for various galaxy samples by 
many authors 
\citep[e.g.][]{Dwek:2007aa,Michaowski:2010aa,Michaowski:2010ab,Hirashita:2011aa,Valiante:2011aa,Kuo:2012aa,Mancini:2015aa,Michaowski:2015aa}. 
Recent experiments by \citet{Rouille:2014ab} showed that accretion 
actually occurs in cold environments.

Less models with a shorter SF time-scale 
($\tau_\mathrm{SF}=5\times 10^8$ yr) reproduce 
the observation extinction curve for all three galaxies than those with 
a longer $\tau_\mathrm{SF}$, which indicates that 
the MW, LMC and SMC have built up stars mildly on time-scales
longer than 1 Gyr, and also implies that
extinction curves might be influenced by the star formation history.
Such a long $\tau_\mathrm{SF}$ is consistent with the pictures adopted 
in other chemical evolution models 
\citep[e.g.][]{Bekki:2015aa}.
For shattering and SN destruction, the parameter 
values we chose are equally favored by the observed extinction curves; 
in other words, it is difficult to constrain these parameters by the 
extinction curves.
This is probably because the small and large grain abundances are
overwhelmed by accretion and coagulation, respectively, with
minor effects of the other processes.

\begin{table*}
\begin{center}
\caption{Fraction of calculated extinction curves that satisfies $\Delta^2 \leq 8$
and minimum $\Delta^2$ under a given value of each fixed parameter.}
\label{table:good_set}
\begin{tabular}{ccc|ccc|ccc|ccc|ccc}
\hline
\multicolumn{15}{c}{Milky Way} \\
\hline
$\tau_\mathrm{SF}$ (yr) & \% & $\Delta^2_\mathrm{min}$ & $\beta_\mathrm{sh,0}$ & \% & $\Delta^2_\mathrm{min}$ & $\beta_\mathrm{co,0}$ & \% & $\Delta^2_\mathrm{min}$ & $\tau_\mathrm{cl}$ (yr) & \% & $\Delta^2_\mathrm{min}$ & $\beta_\mathrm{SN}$ & \% & $\Delta^2_\mathrm{min}$ \\
\hline
$5\times10^8$ & 19 & 0.29 & $5\times10^3$ & 33 & 0.50 & $5\times10^4$ & 52 & 0.50 & $10^6$ & 21 & 3.26 & 4.83 & 41 & 0.29  \\
$5\times10^9$   & 40 & 0.12 & $5\times10^2$ & 46 & 0.12 & $5\times10^3$ & 69 & 0.12 & $10^7$ & 42 & 0.13 & 9.65 & 40 & 0.28  \\
$5\times10^{10}$& 47 & 0.12 & $5\times10^1$ & 40 & 0.29 & $5\times10^2$ & 40 & 0.29 & $10^8$ & 42 & 0.12 & 19.3 & 25 & 0.12  \\ 
 &  &     & $5\times10^0$ & 26 & 0.56 & $5\times10^1$   & 4 & 6.85 & & & & & &\\
 &  &     & $5\times10^{-1}$  & 19 & 0.74 & $5\times10^0$   & 0 & 8.74 & & & & & &\\
\hline
\multicolumn{15}{c}{LMC} \\
\hline
$5\times10^8$ & 15 & 0.66 & $5\times10^3$ & 56 & 1.17 & $5\times10^4$ & 19 & 4.44  & $10^6$ & 5 & 2.56 & 4.83 & 31 & 1.17  \\
$5\times10^9$   & 28 & 0.48 & $5\times10^2$ & 33 & 0.48 & $5\times10^3$ & 54 & 0.48  & $10^7$ & 38 & 0.48 & 9.65 & 33 & 0.66  \\
$5\times10^{10}$& 38 & 0.48 & $5\times10^1$ & 22 & 0.80 & $5\times10^2$ & 40 & 0.66  & $10^8$ & 38 & 0.66 & 19.3 & 17 & 0.48  \\ 
 &  &     & $5\times10^0$   & 20 & 0.66 & $5\times10^1$   & 0 & 11.8 & & & & & & \\
 &  &     & $5\times10^{-1}$  & 15 & 0.66 & $5\times10^0$   & 0 & 32.7 & & & & & & \\
\hline
\multicolumn{15}{c}{SMC} \\
\hline
$5\times10^8$ & 4 & 6.35  & $5\times10^3$ & 19 & 2.34 & $5\times10^4$ & 0 & 16.9 & $10^6$ & 0 & 21.0 & 4.83 & 5 & 4.92  \\
$5\times10^9$ & 9 & 4.92  & $5\times10^2$ & 11 & 4.92 & $5\times10^3$ & 15 & 2.34 & $10^7$ & 2 & 5.20 & 9.65 & 12 & 2.34  \\
$5\times10^{10}$& 12&2.34 & $5\times10^1$ & 6 & 6.35 & $5\times10^2$  & 15 & 4.92 & $10^8$ & 22 & 2.34 & 19.3 & 7 & 2.83  \\ 
 &  &     & $5\times10^0$   & 6 & 6.79 & $5\times10^1$  & 0 & 9.22 & & & & & & \\
 &  &     & $5\times10^{-1}$  & 4 & 6.97 & $5\times10^0$  & 0 & 9.93 & & & & & & \\
\hline
\end{tabular}
\end{center}
Note: For the graphite fraction and $\alpha_\mathrm{C}$, we adopt the values in Table \ref{table:different_model}.
\end{table*}%

\begin{figure*}
\begin{center}
\includegraphics[width=\textwidth]{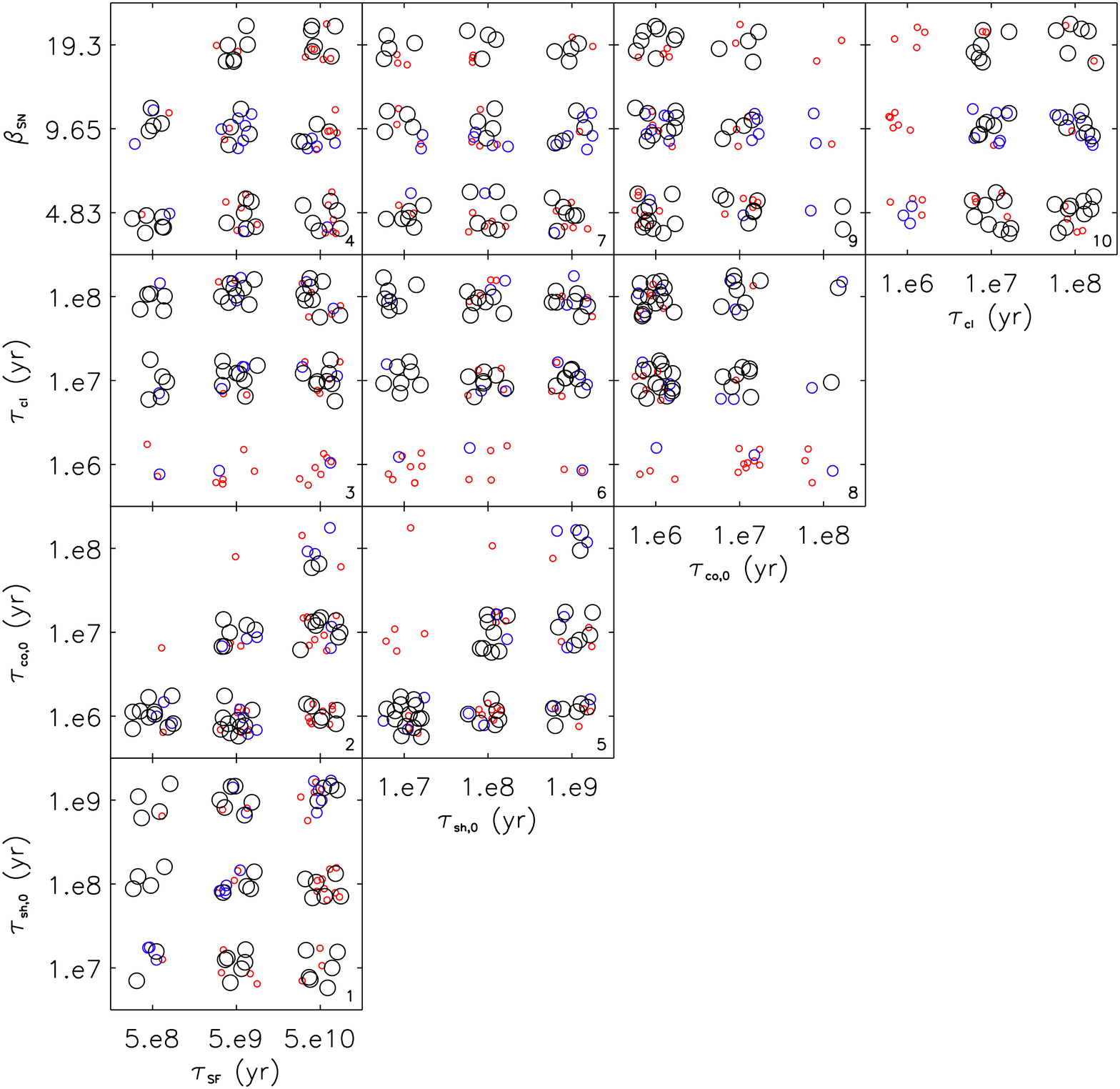}
\caption{
Correlation plots between two parameters chosen in
each panel for the MW case.
Each parameter has three values; thus, each panel is composed of
9 grid points. Since the other three parameters are changed, we have
27 cases for each grid point. Among these 27 cases, we only show the
cases with $\Delta^2 \le 8$, each circle representing one case.
To avoid overlaps of circles, we add a random offset for each circle.
Symbol sizes correspond to values of $\Delta^2$: large black circles are
for $\Delta^2 \le 2$, 
medium blue circles are for $2< \Delta^2 \le 5$, and small red circles are for $5< \Delta^2 \le 8$.
A grid point with many circles means that we find many cases that satisfy
$\Delta^2\le 8$: thus, we regard the grid points with many large points
are favored by the observed MW extinction curve. 
Panels showing $\tau_\mathrm{co,0}$ indicate that strong coagulation
($\tau_\mathrm{co,0} \le 10^7$ yr)  fits the observed data.
Strong accretion ($\tau_\mathrm{cl} \ge 10^7$) is also preferred.
The relations between $\tau_\mathrm{co,0}$ and $\tau_\mathrm{SF}$
and between $\tau_\mathrm{co,0}$ and $\tau_\mathrm{sh,0}$ show
degeneracy in the diagonal direction.
There are no clear constraints on $\tau_\mathrm{sh}$ and $\beta_\mathrm{SN}$.
}
\label{Fig:two parameters MW}
\end{center}
\end{figure*}%

To illustrate our results further, we make correlation plots
between two parameters
in Fig.~\ref{Fig:two parameters MW}.
Each panel shows the relation between two parameters.
For example, the upper right panel shows the constraints on the
$\tau_\mathrm{cl}$--$\beta_\mathrm{SN}$ plane. Since we examine
three values for each parameter, we have nine cases
(nine grid points) in each
panel. Each grid point shows one case with $\Delta^2\leq 8$ with
the two parameters fixed and the other three parameters moved
(so there are 27 cases for each point); thus,
if there are many points plotted, that combination of two parameter values
has many cases that predict extinction curves close to the MW
extinction curve.
Larger symbols mean smaller values of $\Delta^2$ 
(i.e.\ the calculated extinction curve is
closer to the observed one if the symbol size is larger).
For example, in the upper right panel, small $\tau_\mathrm{cl}$ values
(strong grain growth by accretion) is preferred by the Milky Way extinction
curve, while $\beta_\mathrm{SN}$ is not constrained well (i.e.\
all the three values of $\beta_\mathrm{SN}$ are equally preferred).
There are preferred trends for $\tau_\mathrm{co}$ and $\tau_\mathrm{cl}$
in such a way that strong coagulation and accretion accommodate
more extinction curves close to the observed one.
We observe almost no preference for the choice of
$\tau_\mathrm{sh}$ and $\beta_\mathrm{SN}$. A
longer $\tau_\mathrm{SF}$ (mild star formation) is slightly favored than
a shorter $\tau_\mathrm{SF}$ (burst-like star formation).
All these trends support the conclusions derived above as to
Table~\ref{table:good_set}. 
Fig.~\ref{Fig:two parameters MW} also shows a diagonal trend
in the $\tau_\mathrm{co}$--$\tau_\mathrm{SF}$ and
$\tau_\mathrm{co}$--$\tau_\mathrm{sh}$ diagrams, which means that
similar extinction curves are produced with similar values for
ratio $\tau_\mathrm{co}/\tau_\mathrm{SF}$ or
$\tau_\mathrm{co}/\tau_\mathrm{sh}$.
This is because the coagulation time-scale relative to the star formation
time-scale is the real
parameter that determines the grain size distribution,
and the final small-to-large grain abundance ratio is determined by the
balance between coagulation and shattering (see also H15).
We obtained similar trends for preferred parameter ranges for the SMC
and LMC as shown in Appendix~\ref{sec:appendix}.

\subsection{Tuning for the SMC and LMC}

To fit the observed SMC and LMC extinction curves, we introduced
AC for carbonaceous dust and adopted a stronger SN destruction 
efficiency for small carbonaceous dust ($\alpha_\mathrm{C}$) 
(Sections \ref{subsec:smc} and \ref{subsec:lmc}).
The key differences between the MW and SMC/LMC extinction curves are
the strength of 2175 \AA\ bump and the UV slope. Although the 
steep slope can be produced within the 
parameter ranges that we adopted, the lack of the 2175 \AA\ bump feature
cannot be explained as long as we adopt graphite for the carbonaceous
component. We have also shown that, even if we impose a stronger
destruction for carbonaceous dust than for silicate, the
small carbonaceous grains are never eliminated because
accretion is efficient enough to raise the abundance of small
carbonaceous dust grains. Therefore, it was necessary to
introduce AC instead of graphite for the purpose of explaining 
the SMC and LMC extinction
curves.

It may be natural to assume an enhanced destruction efficiency
also for silicate as well as carbonaceous dust. However, if we
adopt $\alpha_\mathrm{Si} = \alpha_\mathrm{C} = 0.1$,
the abundance ratio of carbonaceous dust to silicate 
does not change from the case of $\alpha_\mathrm{Si} = \alpha_\mathrm{C} = 1$, 
so that the feature caused by carbonaceous dust still remains prominently.
Therefore, enhancing the destruction of small carbonaceous grains
compared with that of small silicates is essential in explaining
the extinction curves which lack prominent carbonaceous 
dust features.

\citet{Bekki:2015aa} presented a different way of explaining the lack of
the 2175 \AA\ bump feature for the SMC extinction curve. In their models,
a strong dust wind is assumed to be associated with the starburst events.
Assuming that small graphite grains are easily removed by outflows,
they indeed succeeded in reproducing the SMC and LMC extinction curves
at the present age. Although their scenario is plausible, we have shown
in this paper that modifications of the SN destruction
efficiency and the introduction of AC without any dust outflow
can also explain the SMC/LMC extinction curves.

Assuming AC to be the dominant carbonaceous dust component 
in the Magellanic Clouds may be observationally supported.
It is indicated that the far-infrared SED of the LMC is produced 
by AC and silicate better than the standard graphite and silicate 
dust model \citep{Meixner:2010aa}.
\citet{Jones:2011aa} mentioned that the lifetime of silicate is longer 
than that of AC, which would justify our different SN destruction rates 
between small silicate and AC grains. In this context, the small 
$\alpha_\mathrm{C}$ may be the consequence of introducing an AC species.

\citet{Pei:1992aa}, \citet{Weingartner:2001aa} and \citet{Li:2006aa}
suggested a very small graphite-to-silicate mass ratio to fit the bumpless
SMC extinction curve. Their models are in line with our conclusion
that an enhanced destruction efficiency of carbonaceous dust relative to silicate
better reproduces the SMC extinction curve. However, in our model,
the abundance of small carbonaceous grain is inevitably kept to a
level at which the 2175 \AA\ bump still appears. thus, we needed to 
introduce another carbonaceous species (i.e.\ AC) which does not 
have such a feature.

\subsection{Can we explain all with the same parameter set?}

Using our models, we have made an attempt to explain all the well 
studied extinction curves in the local Universe. Here, we examine 
a possibility that the three extinction curves are understood as 
a single evolutionary sequence.

\begin{figure}
\begin{center}
\includegraphics[width=0.45\textwidth]{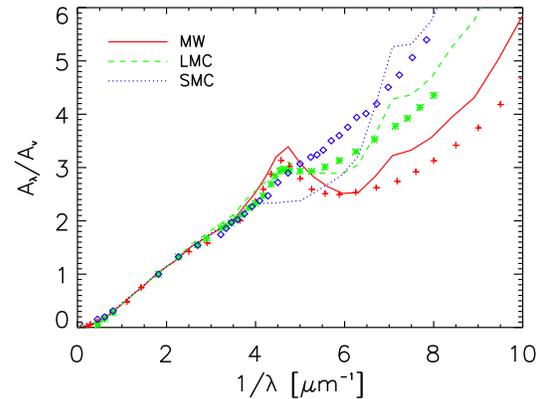}
\caption{Metallicity sequence of extinction curves at 0.2, 0.5, and 1 $Z_{\solar}$ calculated by a single common parameter set (the fiducial parameters in Table \ref{table:parameters}), compared with the SMC, LMC, and MW extinction curves, respectively. For $\alpha_\mathrm{C}$ and the carbonaceous material, we adopt the models in Table \ref{table:different_model}. Cross, star and square symbols represent the observed mean extinction curves of the MW, the LMC and the SMC taken from \citet{Pei:1992aa}. Solid, dashed, and dotted lines are the modeled extinction curves for the MW, the LMC and the SMC.}
\label{Fig:evolution_sequence}
\end{center}
\end{figure}

As shown above, the SMC extinction curve has the severest constraint
on parameters, with only 20 parameter sets fitting
the observational data with $\Delta^2 \leq 8$. Using these 20
parameter sets, we predict the MW and LMC extinction curves
to see if they fit these extinction curves as well.
We adopt the tuned graphite fraction and $\alpha_\mathrm{C}$ 
listed in Table \ref{table:different_model}. As a consequence,
there are 11 parameter sets (5 per sent of all the 243 sets) 
fitting extinction curves of the MW, LMC and SMC satisfactorily
($\Delta^2 \leq 8$) at appropriate metallicities
$Z =$ 1.0, 0.5 and 0.2 $Z_{\odot}$, respectively.
Fig.\ \ref{Fig:evolution_sequence} shows an example out of
the 11 parameter sets.
Although we do not intend to argue that all the galaxies 
have the same dust processing time-scales, we emphasize that 
we have succeeded in explaining the three well known extinction curves with 
a single dust evolution framework.

\subsection{Extinction curves in other galaxies}

We have concentrated on the local extinction curves which
are studied in details. Moreover, since the extinction curves
are derived for individual lines of sight toward single
stars, extinction curves likely reflect the dust properties
in the ISM with minor contribution from radiative transfer
effects. For example, the Calzetti extinction curve
\citep{Calzetti:1994aa} represents an effective
extinction as a result of radiative transfer
(absorption and scattering) effects of light emitted from
multiple stellar populations 
\citep[see also][]{Inoue:2005aa}.
Extinction curves derived for galaxies whose stars cannot
be resolved have in principle the same problem.
Careful radiative transfer modeling would be necessary 
in such a case and is left for future work.

In the distant Universe, quasars and gamma-ray bursts
are suitable to derive the extinction curves in their
host galaxies because of their simple power-law-like
spectral shapes
\citep{Maiolino:2004aa,Stratta:2007aa,Li:2008ab,Eliasdottir:2009aa,
Gallerani:2010aa,Perley:2010aa,Zafar:2010aa,Zafar:2011aa,Zafar:2015aa,Schady:2012aa}.
Moreover, they are very bright point sources.
Therefore, their extinction curves likely reflect the
intrinsic dust properties in the host galaxies, although we should keep in
mind that radiative transfer effects, especially scattering,
skew the shape of the extinction curve if we deal with
highly extinguished objects \citep[e.g.][]{Scicluna:2015aa}.
A successful fit to the extinction curve of a quasar
at $z=6.2$ is given by
\citet{Nozawa:2015aa}, who used their evolution model of
grain size distribution.
Using our models, we will further
examine the diverse types of
extinction curves actually observed in those high-redshift
objects in the future.

\section{conclusion}

Using the dust evolution model developed in our previous work,
we investigated if the extinction curves in the local
galaxies can be explained by well known dust enrichment
processes. In order to make a parameter survey possible,
we adopted the two-size approximation in which the full range
of grain sizes is represented by the small and large sizes divided at $a \sim$ 0.03 $\mu$m.
We considered two dust species, carbonaceous 
dust and silicate, and calculated the evolution of the large 
and small grain abundances for each species. For the
processes driving the evolution of dust, we considered
dust condensation in stellar ejecta,
destruction in SN shocks, fragmentation by shattering and growth
by coagulation and accretion.
To calculate the MW extinction curves, we adopted `standard' dust 
composition, silicate and graphite. Since the 
model is computationally light, we surveyed the reasonable ranges 
for the time-scale of each process and consequently found 37 per cent
of the parameter 
sets fitting the observed MW extinction curve. This means that
the known processes driving the evolution of dust successfully
reproduces the dust properties consistent with the MW extinction
curve. We also showed that the extinction curves 
are sensitive to the star formation, coagulation and accretion processes.

The same dust evolution model failed to reproduce the SMC 
extinction curves.
This is because the 2175 \AA\ bump feature remains for any 
parameter set. We proposed two possible modifications: using AC instead of 
graphite for carbonaceous dust and/or adjusting SN destruction 
efficiency for small carbonaceous grains.
Using AC for carbonaceous dust or adopting a higher SN
destruction rate for small carbonaceous dust
($\alpha_\mathrm{C} = 0.1$), we found some cases where the extinction
curve fits the observed SMC extinction curve.
Adopting both modifications at the same time, the fitting improved;
in particular, since small carbonaceous dust inevitably remains
because of accretion (recall that accretion selectively occurs for
small grains), assuming graphite always leads to a significant
2175 \AA\ feature. Thus, considering grain species other than graphite
is essential in explaining the SMC extinction curve.
We also confirmed that the LMC extinction curve is explained by
an intermediate value of $\alpha_\mathrm{C}(=0.2)$ and
a mixture of AC and graphite for the carbonaceous component.

By analyzing the favored parameter sets by each extinction curve,
we obtained constraint on coagulation and accretion processes.
Overall, strong coagulation
($\beta_\mathrm{co,0}\gtrsim 500$, which means that the
coagulation time-scale under the MW dust-to-gas ratio
is shorter than $~1/500$ of the star-formation time-scale)
is favored, especially for the MW.
Efficient grain growth by accretion under long cloud lifetimes 
($\tau_\mathrm{cl}\sim 10^7$--$10^8$ yr) fits
the observed extinction curves in all the three galaxies,
especially the SMC.
Longer star formation time-scales than $10^9$ yr is also preferred,
which is consistent with the picture that those three galaxies have
been built up on time-scales comparable to the cosmic age.
Other processes are not constrained strongly 
by observed extinction curves, which implies that 
they do not have as large a imprint in the dust evolution
as accretion and coagulation. Finally, we also presented a possibility of
explaining all the three extinction curves as a sequence of metallicity
evolution.

\section*{Acknowledgments}

We thank the anonymous referee for useful comments.
We are grateful to Typhoon Lee and You-Hua Chu for their support of KCH's PhD program.
We also thank Kentaro Nagamine, Shohei Aoyama, Ikkoh Shimizu and Peter Scicluna for
stimulating discussions.
HH is supported by the Ministry of Science and Technology
(MoST) grant 105-2112-M-001-027-MY3.
MJM acknowledges the support of the UK Science and Technology
Facilities Council, the Royal Society of Edinburgh International
Exchange Program and the hospitality of the Academia Sinica
Institute of Astronomy and Astrophysics.

\appendix
\section{Correlation plots for the LMC and SMC}
\label{sec:appendix}

We show the correlation plots for the LMC and SMC.
They are produced in the same way as the MW case
in Section \ref{subsec:constraint} but using the
tuned parameters listed in Table \ref{table:different_model}.

\begin{figure*}
\begin{center}
\includegraphics[width=\textwidth]{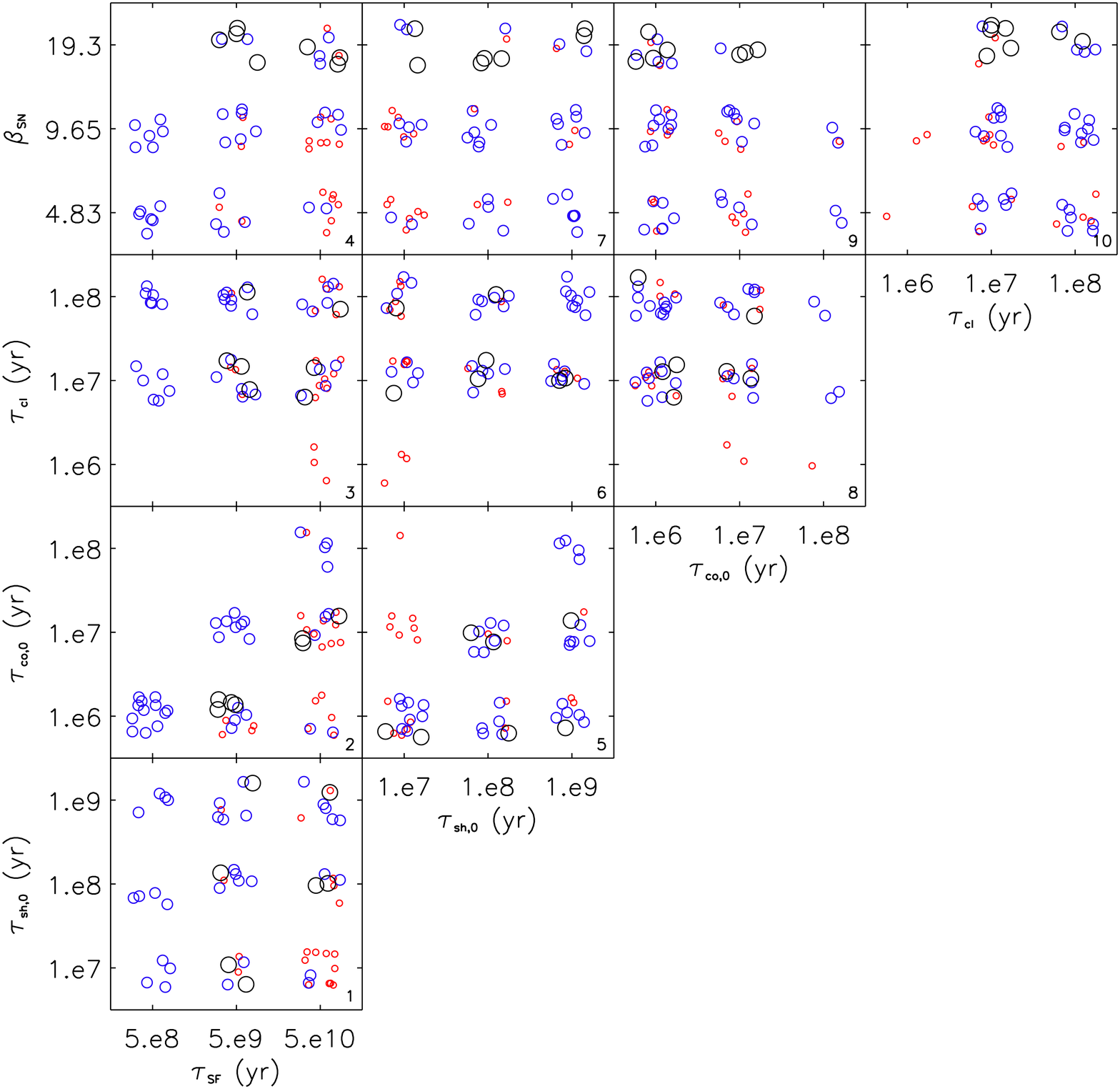}
\caption{Same as Fig. \ref{Fig:two parameters MW} but for the LMC.
}
\label{Fig:two parameters LMC}
\end{center}
\end{figure*}%

\begin{figure*}
\begin{center}
\includegraphics[width=\textwidth]{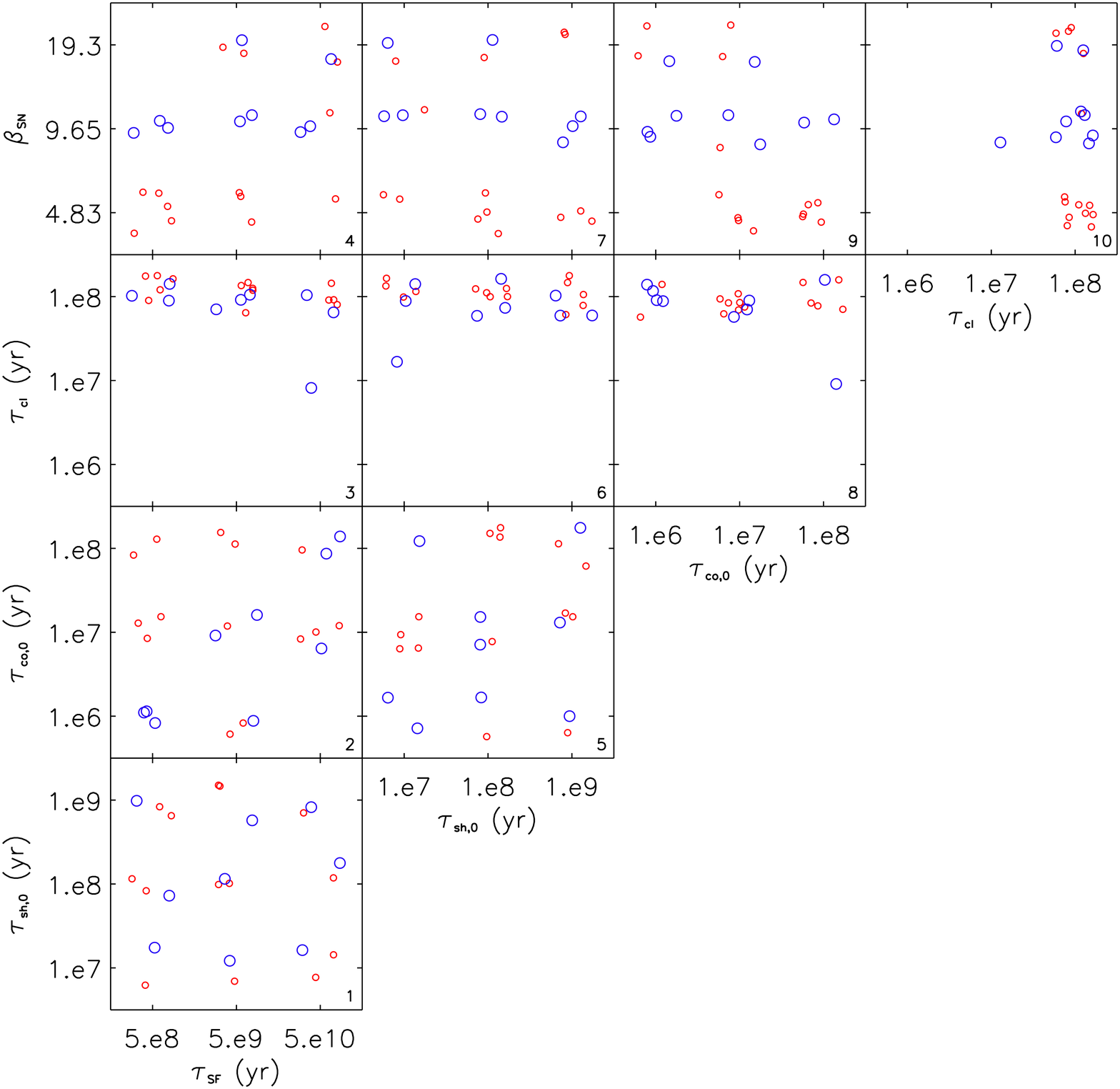}
\caption{Same as Fig. \ref{Fig:two parameters MW} but for the SMC.
}
\label{Fig:two parameters SMC}
\end{center}
\end{figure*}%

\bibliography{YMO_pasj}

\end{document}